\documentclass[%
 reprint,
 amsmath,amssymb,
 aps,
floatfix,
]{revtex4-2}
\usepackage{graphicx}
\usepackage{dcolumn}
\usepackage{bm}

\usepackage{nicefrac}
\usepackage[euler]{textgreek}	
\usepackage[normalem]{ulem}	
\usepackage{physics}		

\usepackage{hyperref}

\DeclareMathOperator{\e}{e}
\newcommand{\mrm}{\mathrm}
\newcommand{\mcal}{\mathcal}

\newcommand{\kT}{k_{\mathrm{B}}T}
\newcommand{\ti}{\textit}

\newcommand{\trelo}{\tau_{\mrm{r}}}

\newcommand{\dgr}{\delta_{\mrm{g}}}

\newcommand{\ts}{t_{\mrm{s}}}
\newcommand{\fs}{f_{\mrm{s}}}

\newcommand{\xnr}{x_{n}^{\mrm{r}}}
\newcommand{\xnnr}{x_{n+1}^{\mrm{r}}}
\newcommand{\xnpr}{x_{n^{+}}^{\mrm{r}}}

\newcommand{\pr}[1]{\left(#1\right)} 
\newcommand{\sr}[1]{\left[#1\right]} 

\newcommand{\xT}{X_\mrm{T}}
\newcommand{\xR}{X_\mrm{R}}
\newcommand{\SI}{\textit{Appendix}}

\newcommand{\beginsupplement}{%
        \setcounter{table}{0}
        \renewcommand{\thetable}{\arabic{table}}%
        \setcounter{figure}{0}
       \renewcommand{\thefigure}{S\arabic{figure}}
        }

\begin{document}

\preprint{APS/123-QED}

\title{Maximizing power and velocity of an information engine}

\author{Tushar K. Saha, Joseph N. E. Lucero, Jannik Ehrich}
\author{ David A. Sivak}
\homepage{email: dsivak@sfu.ca}
\author{John Bechhoefer}
\altaffiliation{email: johnb@sfu.ca}

\affiliation{
Dept. of Physics, Simon Fraser University, Burnaby, British Columbia, V5A 1S6, Canada
}

\date{\today}

\begin{abstract}
Information-driven engines that rectify thermal fluctuations are a modern realization of the Maxwell-demon thought experiment. We introduce a simple design based on a heavy colloidal particle, held by an optical trap and immersed in water. Using a carefully designed feedback loop, our experimental realization of an ``information ratchet'' takes advantage of favorable ``up'' fluctuations to lift a weight against gravity, storing potential energy without doing external work. By optimizing the ratchet design for performance via a simple theory, we find that the rate of work storage and velocity of directed motion is limited only by the physical parameters of the engine: the size of the particle, stiffness of the ratchet spring, friction produced by the motion, and temperature of the surrounding medium. Notably, because performance saturates with increasing frequency of observations, the measurement process is not a limiting factor. The extracted power and velocity are at least an order of magnitude higher than in previously reported engines. 

\end{abstract}

\maketitle


Over 150 years ago, Maxwell proposed a thought experiment to sharpen understanding of the second law of thermodynamics \cite{maxwell1867}. 
He envisioned a ``neat-fingered being'' that could sort fast and slow molecules to create a temperature difference between two chambers, thereby converting the energy of a heat bath into a form that could be used to do work. In modern terms, Maxwell's thought experiment was the first example of an \ti{information engine}.  In 1929, Leo Szilard proposed a simpler variant consisting of a single gas molecule in a chamber, partitioned by a wall~\cite{szilard1929,szilard1929a}. If the particle is observed in the left half, the demon attaches a mass raised by motion to the right, and vice versa.  Then, an isothermal expansion of the chamber raises the mass and stores potential energy. When run cyclically, the engine converts information about the state of the molecule into gravitational potential of a raised mass, seemingly without doing any work to lift the mass. This apparent violation of the second law of thermodynamics was later resolved by considering costs associated with the processing of information~\cite{landauer1961, bennett1982}, leading to a clearer understanding of the thermodynamics of information~\cite{mandal2012,schmitt2015,parrondo2015}.

Recent advances in technology and theoretical developments in stochastic thermodynamics~\cite{sekimoto1997,sekimoto2010,seifert2012,vandenbroeck2015} have made it possible to experimentally realize information engines, based on the ideas of Maxwell and Szilard~\cite{toyabe2010, camati2016, koski2015}. They have been used to evaluate the Landauer cost of manipulating the associated measurement-memory device~\cite{berut2012, jun2014,  koski2014prl, hong2016} and to explore the efficiency of information-to-work conversion~\cite{ribezzi2019,admon2018,paneru2018pre, paneru2020Natcomm}. 

Here, we create and study the performance of a \ti{useful} information engine that not only extracts energy from heat but also \ti{stores energy} by raising a weight, as initially imagined by Szilard. The ``fuel'' for the motor is the information gathered from favorable system fluctuations.

\begin{figure}[ht]
\centering
\includegraphics[width=\linewidth]{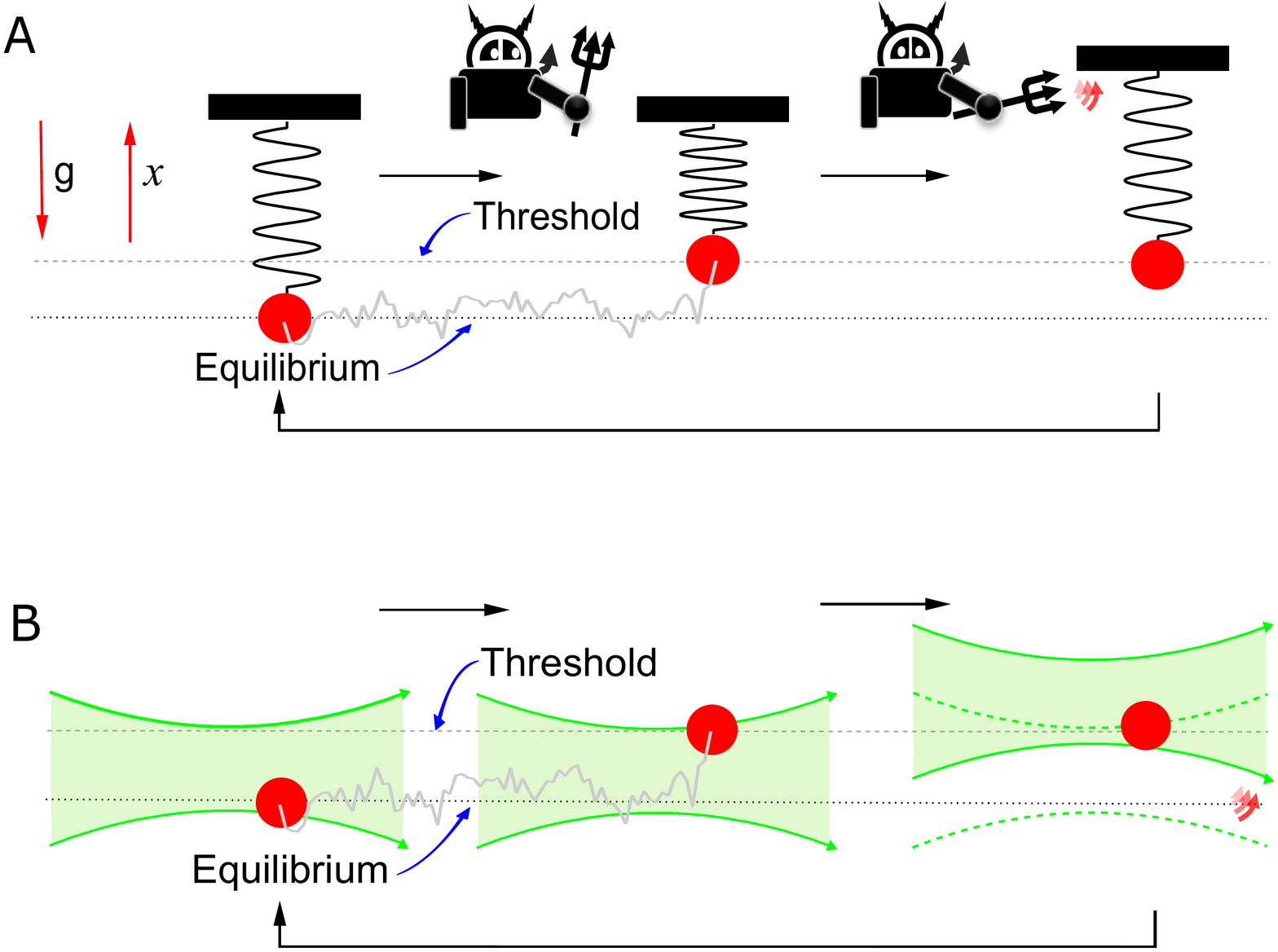}
\caption{Schematic of the information engine.  
(A) Ratcheted spring-mass system under gravity.  
(B) Experimental realization using horizontal optical tweezers in a vertical gravitational field.  
Feedback operations on the right side in (A) and (B) are indicated by the small red ``swoosh'' arrows.}
\label{fig:Schematic}
\end{figure}
Our information engine consists of an optically trapped, micron-scale bead in water. The laser beam of the trap is horizontal, perpendicular to the vertical gravitational axis. The optical tweezers create a harmonic potential, where the bead fluctuates about an equilibrium that is lower because of the bead's weight (Fig.~\ref{fig:Schematic}\ti{B}). The motion of the heavy bead can be modeled by a simple spring-mass system (Fig.~\ref{fig:Schematic}\ti{A}). The demon monitors the position of the mass and, when the mass fluctuates beyond a predefined threshold, raises the position of the spring anchor (top bar). Repeating the process, the mass is raised by exploiting favorable ``up'' fluctuations arising from thermal noise in the medium. 

The experimental setup is similar to \cite{paneru2018prl}, but here we store the extracted work in a reservoir. The ability to ``spend'' stored work on demand and for varying purposes greatly increases the utility of the engine. A previous experimental system introduced by Admon et al.~\cite{admon2018} also stored work, but its design was based on a repulsive potential, which meant that the motor was always powered by a combination of external mechanical work and information.  Here, with a design based on a trap potential having a local minimum, we ensure that no external work is done on the bead, which simplifies the physical picture.
In our study of this new information engine, we focus on understanding and then optimizing its performance:  How much can it lift? How fast can it go? More precisely, What is the upper bound to the rate of gravitational energy storage and to the directed velocity? We reason that the value of the \ti{function} of a motor can greatly exceed the cost of running it. For example, in biological applications such as chemotaxis, the metabolic costs of running cellular machinery (including information-processing costs) are usually unimportant compared to the benefit gained by the ability to move toward a new food source or away from a predator~\cite{berg2004}. We thus seek to maximize performance, independent of the energy required. As we will show, there is a maximum achievable energy-storage rate and a maximum achievable directed velocity, even when the signal-to-noise ratio of the measuring system is arbitrarily high (with correspondingly high costs for information processing); knowing the maximum level of performance independent of information costs can provide a benchmark to evaluate trade-offs between performance and operational costs. We will also show that the performance of an information engine is limited by its material parameters. In our case, these parameters include trap stiffness and bead size, and we provide a systematic method of choosing their values to maximize the desired performance measure.

\section*{Theory}
\subsection*{Equation of Motion}

The dynamics of an optically trapped bead are well described by an overdamped Langevin equation,
\begin{align}
    \gamma \dot{x}(t) =\underbrace{- \kappa \left(x(t)-\lambda(t)\right)}_{\text{restoring force}}-\underbrace{mg}_{\text{grav.~force}} +\underbrace{\sqrt{2k_{\rm{B}}T\gamma}\ \nu(t)}_{\text{thermal noise}}\ ,
\label{eq:EQM_full}
\end{align}
where $x(t)$ denotes the position of a bead of radius $r$ at time $t$, $\lambda(t)$ the center of the trap, $\kappa$ the trap stiffness, $\gamma$ the friction coefficient, $g$ the gravitational acceleration, and $\nu(t)$ represents Gaussian white noise with zero mean and $\langle \nu(t) \, \nu(t') \rangle = \delta(t-t')$.  The effective mass $m = (4/3)\pi r^3 \Delta \rho$ of the bead depends on the relative density $\Delta \rho = \rho_\textrm{bead} -\rho_\textrm{medium}$ and accounts for buoyancy. 
Scaling lengths by the equilibrium standard deviation $\sigma = \sqrt{k_{\rm{B}} T/\kappa}$ of the bead position and time by the bead relaxation time $\tau_r = \gamma/\kappa$, the overdamped Langevin equation becomes
\begin{align}
    \dot{x}(t) = - \left[ x(t) - \lambda(t) \right] - \dgr + \sqrt{2}\ \nu(t)\ , 
\label{eq:non-dim_EQM}
\end{align}
where $\dgr \equiv mg/\kappa\sigma$ is a scaled effective mass that measures the sag of the bead due to gravity, relative to the scale of equilibrium fluctuations in the trap.  The bead position is measured at discrete time intervals of $\ts' =  20 $ \textmu s, and the feedback on the trap position is applied after a delay of one time step.  Integrating Eq.~\ref{eq:non-dim_EQM} over one time step gives discrete-time dynamics~\cite{kloeden2013},
\begin{align}
    x_{n+1} = \e^{-\ts} x_n  + \left( 1 -\e^{-\ts} \right) ( \lambda_n - \dgr ) + \sqrt{1 - \e^{-2\ts}}\ \xi_n  \,, 
\label{eq:discrete_EQM}
\end{align}
where $\ts= \ts'/\tau_\mrm{r}$, $x_n \equiv x(n \ts)$ denotes the position at time step number $n$, and $\xi_{n}$ is a Gaussian random variable, with zero mean and unit variance, satisfying $\langle \xi_m \, \xi_n \rangle = \delta_{mn}$.

\begin{figure}
\centering
\includegraphics[width=\linewidth]{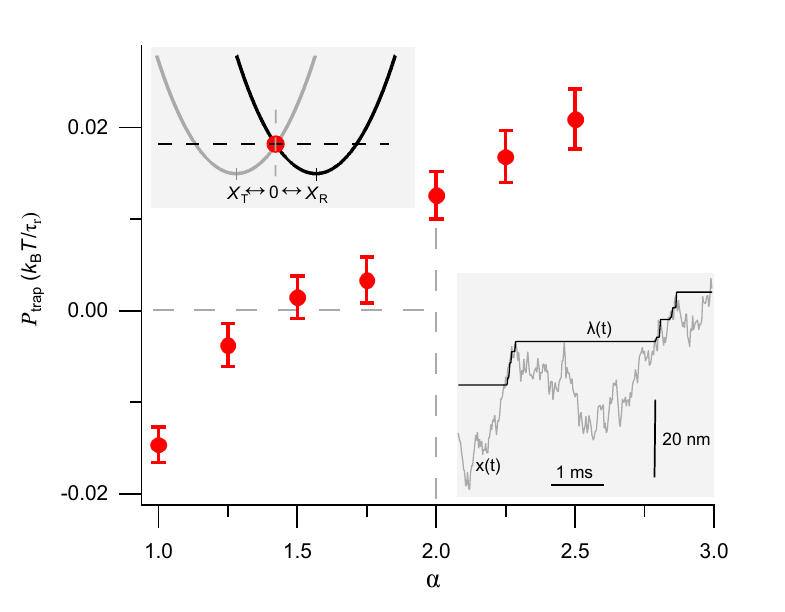}
\caption{
    Zero-work condition defining a pure information engine. 
    Trap power $P_{\mrm{trap}}$ as a function of feedback gain $\alpha$ for fixed threshold $\xT = 0$, scaled relaxation time $\trelo/\ts' = 180$, and scaled effective mass $\dgr = 0.8$. Bottom right inset: Experimental trajectories of the bead ($x(t)$, gray) and trap ($\lambda(t)$, black) during continuous ratcheting. Top left inset: Naive zero-work condition for a harmonic potential is $\alpha=2$, equivalent to $\xR = \xT$. The black curve denotes the trap potential in the current step and gray curve in the previous step.  Error bars here and in other figures are the standard error of the mean (\ti{Appendix}, section \ref{App:M}).} 
\label{fig:Pure_IE}
\end{figure}

The trap position $\lambda_n$ is updated according to a feedback algorithm,
\begin{align}
    \lambda_{n+1} =
    \begin{cases}
         \lambda_n + \alpha (x _{n} -\lambda_{n}),&\quad x_n -\lambda_n > \xT \\
         \lambda_n, & \quad\mrm{otherwise} \,.
    \end{cases}
\label{eq:feedback-alg}
\end{align}
Here, $\xT$ is the threshold, and $\alpha$ is the feedback gain.  Figure~\ref{fig:Pure_IE} 
(bottom-right inset) shows example time series of the upward motion $x(t)$ of the mass and $\lambda(t)$ of the trap.

For an instantaneous measurement and shift of the trap center, Fig.~\ref{fig:Pure_IE}  (top-left inset) shows that choosing $\alpha = 2$ would impose a zero-work condition
(cf.~\SI, section \ref{App:B}), where the stored potential energy results solely from the conversion of the information about the bead position; i.e., the work done by the trap is set to zero.  In our experimental apparatus, there is a delay of 20 \textmu s (one time step) arising (mostly) from the acousto-optic deflector (AOD) that controls the position of the trap~\cite{kumar2018nanoscale}.  
During the delay, the bead tends to move back towards the equilibrium, reducing the value of $\alpha$ needed to impose zero work.  Thus, $\alpha$ is set empirically to implement the zero-work condition, which occurs at $\alpha \approx 1.5$ in Fig.~\ref{fig:Pure_IE}. Note that the reset position $\xR$ illustrated in Fig.~\ref{fig:Pure_IE} is related to $\alpha$ by $\xR= (\alpha-1) \xT$.

\subsection*{Energy Storage and Directed Motion}
\label{ssec:energy_cal}

The input work is the change in energy of the bead that occurs when the position $\lambda$ of the trap center is moved. Since the trap center is moved only at the sampling times $\{ t_{n+1} \}$ and since the shift happens at a faster time scale ($< 1$ \textmu s, set by the response of the AOD) than bead motion, the work done at each update $t_{n+1}$ is
\begin{align}
    W_{n+1} = \tfrac{1}{2}\left[\pr{x_{n+1}-\lambda_{n+1}}^{2}-\pr{x_{n+1}-\lambda_{n}}^{2}\right] \,.
\label{eq:discrete_trapwork}
\end{align}
Similarly, the gain in gravitational potential is
\begin{align}
    \Delta U_{n+1} = \dgr \pr{x_{n+1}-x_n} \,.
\label{eq:discrete_grav_energy}
\end{align}
By convention, the trap work is positive if energy flows into the system and negative if it flows out.  

We quantify the performance of the information engine by the (long-time average) directed velocity and stored power, ideally for an infinitely long trajectory. Each trajectory can be viewed as a sequence of independent \ti{ratchet events}, each starting with the particle at position $\lambda-\xR$ inside the trapping potential and ending when the particle fluctuates up and first reaches the position $\lambda + \xT$. The displacement $\Delta x = (\lambda+\xT) - (\lambda-\xR) = \xR+\xT$ is thus fixed for each event, but the time required for event $m$, the \ti{first-passage time} $\tau_\mrm{FP}$, is stochastic~\cite{haenggi1990}. Using the above definitions, we write the velocity

\begin{align}
    v &=\nonumber \lim_{t_\mrm{traj}\to\infty} \frac{X_\mrm{traj}}{t_\mrm{traj}}\\
    &=\nonumber \lim_{N_\mrm{ratch}\to\infty} \frac{\sum_{m = 1}^{N_\mrm{ratch}}\Delta x}{\sum_{m = 1}^{N_\mrm{ratch}}(\tau_\mrm{FP})_m} \\
    &= \frac{\xR+\xT}{\tau_\mrm{MFP}} , 
    \label{eq:vel}
\end{align}
where $X_\mrm{traj}$, $t_\mrm{traj}$ are the total trajectory length and time and $N_\mrm{ratch}$ is the number of ratcheting events. 
We used the law of large numbers to write $\sum_m (\tau_\mrm{FP})_m \to N_\mrm{ratch} \tau_\mrm{MFP}$, with $\tau_{\mrm{MFP}}$ the \ti{mean first-passage time} (MFPT), the average of $\tau_\mrm{FP}$ (cf.~\SI, section \ref{App:J}). 

The corresponding rate of energy extraction (power) is $v \, mg$, or, in scaled units,
\begin{align}
        P = v\ \dgr \,.  
\label{eq:Power_to_velocity}
\end{align}
For each data point, typically 100 repeated trajectories are measured over a fixed distance of 340 nm. The velocity and power are estimated by replacing $X_\mrm{traj}$ and $t_\mrm{traj}$ in Eqs.~\ref{eq:vel} and \ref{eq:Power_to_velocity} with their trajectory averages (\ti{Materials and Methods}).

\subsection*{Predicted Maximum Output Power and Velocity}
\label{ssec:MFPT}

To predict the maximum output power and velocity, we first calculate the MFPT (Fig.~\ref{fig:Pure_IE} top inset). A standard calculation~\cite{haenggi1990,chupeau2020} 
(\SI, section \ref{App:C}) gives, in scaled units,
\begin{align}
    \tau_{\mrm{MFP}}(\xT) = \int_{-\xT}^{\xT} \dd x'\ \e^{V(x')}\int_{-\infty}^{x'} \dd x''\  \e^{-V(x'')}\ ,
\label{eq:FP}
\end{align}
for total potential $V(x) \equiv \frac{1}{2}x^2 + \dgr \,x$\,. Although Eq.~\ref{eq:FP} in general must be solved numerically, a Taylor expansion for small threshold $\xT$ gives, 
\begin{align}
    \tau_{\mrm{MFP}}(\xT) &= \sqrt{2\pi}\ \e^{\dgr^2/2}\ \left[ 1 + \erf\left(\frac{\dgr}{\sqrt{2}}\right)\right]\xT + \mathcal{O}\left(\xT^3\right)\,,
\label{eq:tauMFPT}
\end{align}
with positive higher-order corrections.  

The velocity is then maximized by taking $\xT \to 0$:
\begin{subequations}
    \begin{align}
    v(\xT) &= \frac{2\xT}{\tau_{\mrm{MFP}}(\xT)}\label{eq:analytic_full} \\[3pt]
    &\overset{\xT \to 0}{\longrightarrow} \sqrt{\frac{2}{\pi}}\ \e^{-\dgr^2/2}\ \left[1 + \erf\left(\frac{\dgr}{\sqrt{2}}\right)\right]^{-1} \,.
\label{eq:analytic_scaled}
\end{align}
\end{subequations}
Equation~\ref{eq:analytic_scaled} was derived previously using a different method and in a slightly different context~\cite{park2016}.

In physical units and for large force constants ($\kappa \to \infty$), the velocity and power are
\begin{subequations} 
    \begin{align}
        v' &= \left( \frac{\sigma}{\trelo} \right) v \ \overset{\kappa \to \infty}{\huge \text{$\sim$} } \ 
        \sqrt{\frac{2 k_{\rm B}T}{\pi}} \frac{\sqrt{\kappa}}{\gamma}\ \label{eq:analytic_real_velocity} , \\[3pt]
        P' &= \left( \frac{\kT}{\trelo} \right) P \ \overset{\kappa \to \infty}{\huge \text{$\sim$} } \ \sqrt{\frac{2\kT}{\pi}} \frac{\sqrt{\kappa}}{\gamma} mg \,. \label{eq:real_power_expansion} 
    \end{align}
\label{eq:analytic_real}
\end{subequations}

\section*{Results}

\begin{figure}[t]
    \centering
    \includegraphics[width=\linewidth]{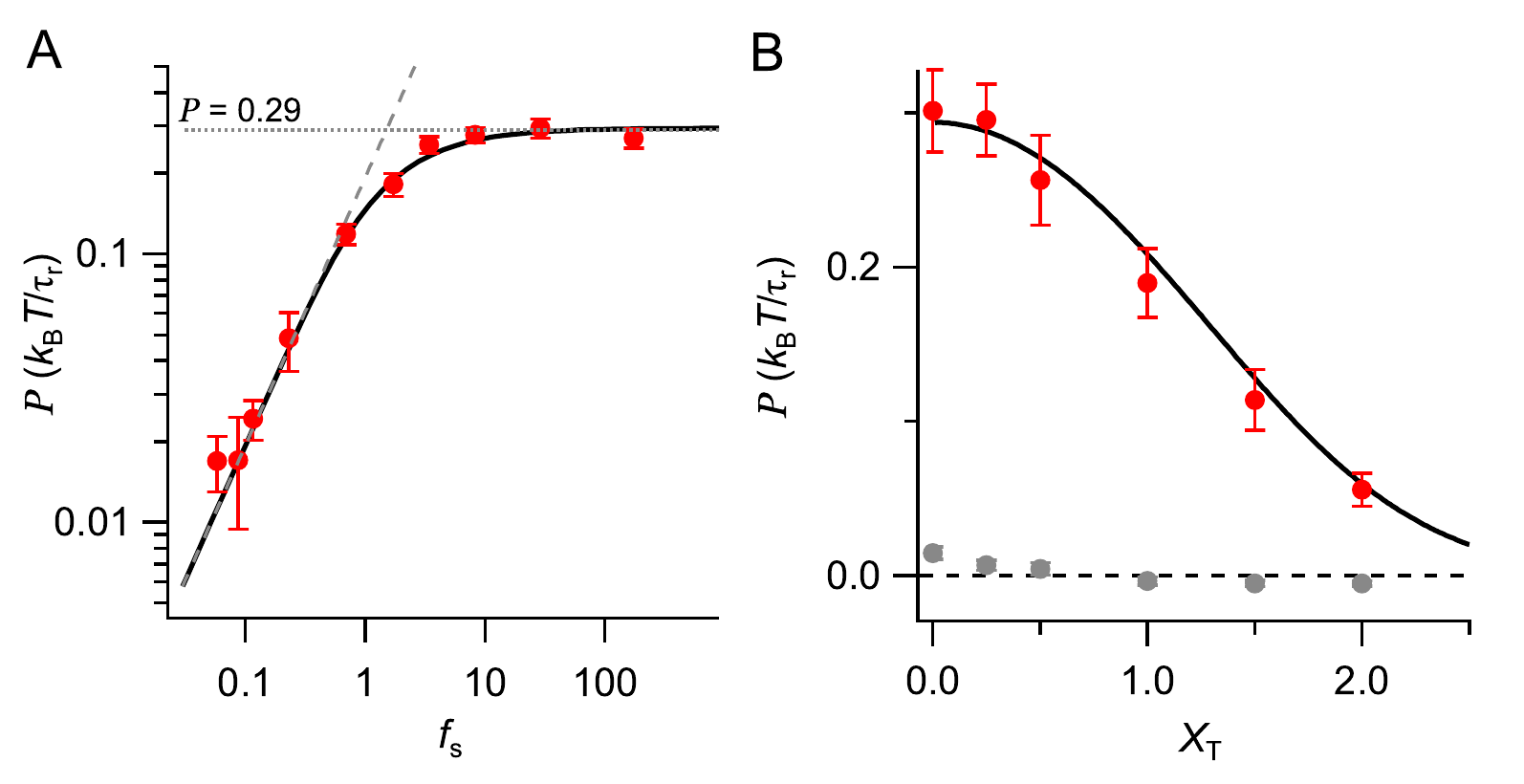}
    \caption{Optimization of ratcheting power. (A) Power as a function of sampling frequency $f_\mrm{s} = \trelo / \ts'$. The black solid curve denotes the semi-analytic results (\SI, section \ref{App:E}) for the same material parameters and $\xT = 0$. The horizontal dotted line indicates the infinite-frequency limit from Eq.~\ref{eq:analytic_real}, and the dashed line denotes the low-frequency limit (\SI, section \ref{App:D}).  
        (B) Power as a function of threshold $\xT$ for fixed $\alpha=1.9$ and sampling frequency of 50 kHz.  The gray markers show that the input trap power is small.  The black curve follows from Eq.~\ref{eq:analytic_full}. Red markers denote experimental values.  For all data, $\dgr = 0.8$.
    }
    \label{fig:optimization}
\end{figure}

To maximize the rate of gravitational-energy extraction (the power), we first studied its dependence on the sampling frequency. Fixing the trap stiffness $\kappa$ and hence the relaxation time $\trelo$, we varied the sampling time $\ts'$.  Figure~\ref{fig:optimization}\ti{A} shows that the power saturates at large sampling frequencies ($\fs = \trelo / \ts' \gg 1$). Thus, making more measurements may not increase the extracted power.  Indeed, measurements faster than the relaxation time $\trelo$ of the bead are correlated and thus provide less information than a single, isolated measurement~\cite{admon2018}. Nonetheless, sampling faster than $\trelo$ reduces the chance of missing a favorable fluctuation that reaches $\xT$. 

At low frequencies, the number of ratchet events is linearly proportional to the sampling frequency.  The gray dotted line in Fig.~\ref{fig:optimization}\ti{A} has slope $\approx 0.19$, which is consistent with a calculation assuming the particle position distribution equilibrates during each interval (\SI, section \ref{App:D}). The solid curve in Fig.~\ref{fig:optimization}\ti{A} is based on semi-analytic calculations (\SI, section \ref{App:E}) that use the measured material parameters and agree well with experiments, with no free parameters. Thus, sampling more slowly than the fluctuation time scale $\trelo$ of the dynamics misses possibly useful fluctuations; sampling more quickly eventually yields diminishing returns.

Having established that the extracted power is maximized for infinite sampling frequency, we henceforth use the fastest feedback time of 20 \textmu s, which typically corresponds to $\fs \gtrsim 100$. Such a sampling frequency is high enough that analytic calculations based on the continuous-sampling limit ($\fs \to \infty$) describe the data well.

We next explored how to set the position threshold $\xT$. This parameter controls the magnitude of the fluctuation that is captured during each ratchet event. The experiments were performed for $\dgr = 0.8$. The feedback gain $\alpha = 1.9$ ensured that the input power was zero for the chosen threshold values, as confirmed by the gray solid markers in Fig.~\ref{fig:optimization}\ti{B}.

Figure~\ref{fig:optimization}\ti{B} shows that the output power, under the constraint of zero input power, is maximized for $\xT \to 0$ (``continuous ratcheting''). The trap position $\lambda(t)$ then either ratchets to accommodate up fluctuations or pauses when the bead fluctuates down, before reaching the threshold again (Fig.~\ref{fig:Pure_IE}, bottom inset). As $\xT$ increases, the fluctuations that take the bead to the threshold become increasingly rare (exponentially in $\xT$), leading to longer wait times between ratchet events; hence, the power tends to zero.  The solid black curve is calculated by numerically integrating Eq.~\ref{eq:FP} to find $\tau_{\mrm{MFP}}(\xT)$.

\begin{figure*}[t]
    \centering
    \includegraphics[width=1\linewidth]{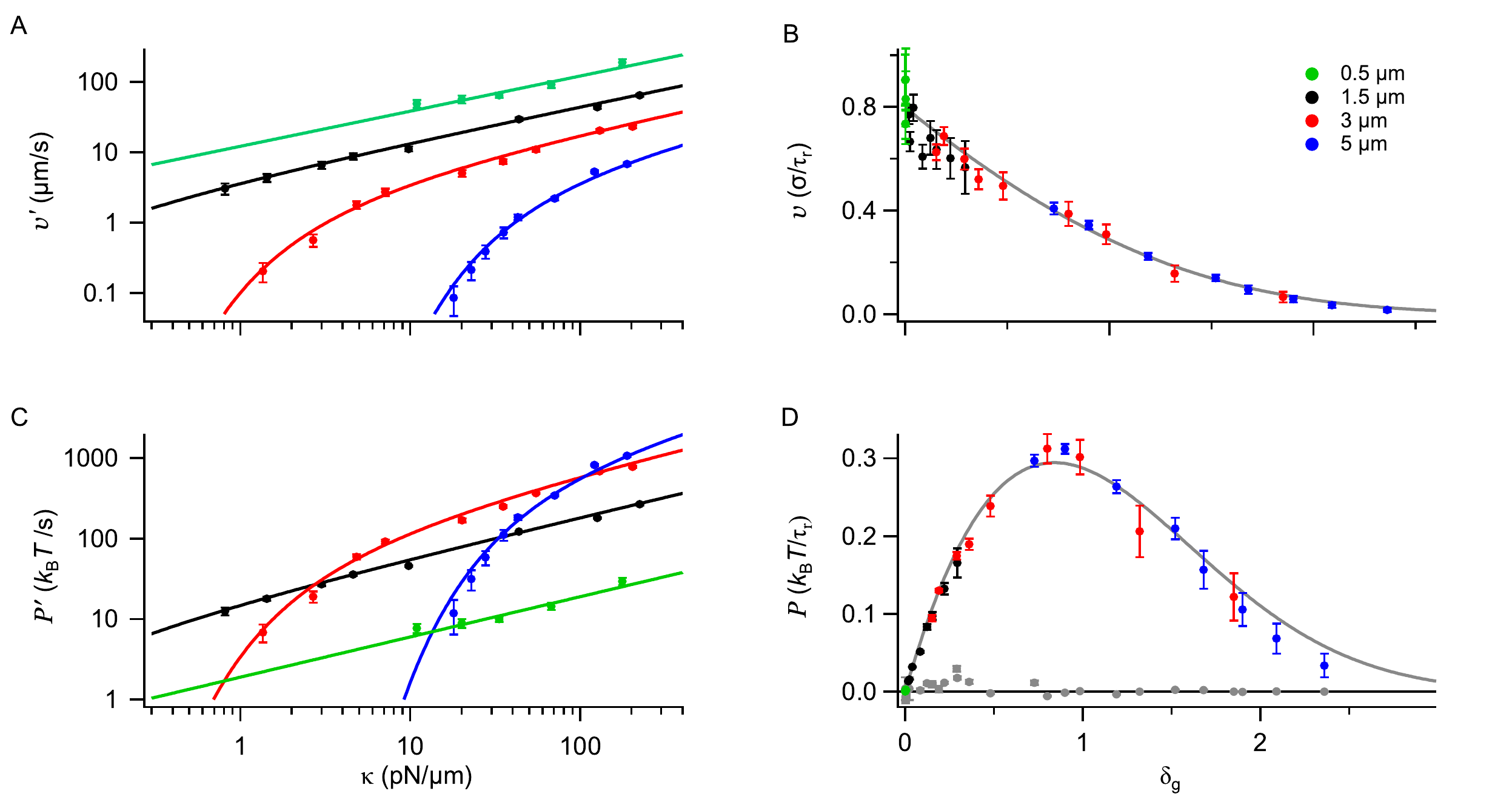}
    \caption{
        Power and velocity for bead diameters 0.5 (green), 1.5 (black), 3 (red), and 5 (blue) \textmu m.  Markers denote experimental data.
        (A) Velocity as a function of trap stiffness $\kappa$. (B) Scaled velocity as a function of scaled effective mass
        $\dgr$.  
        (C) Power as a function of $\kappa$. 
        (D) Scaled power as a function of $\dgr$. Gray markers show that the corresponding $P_{\mrm{trap}}$ values remain small. Solid curves in (A) and (C) are calculated from Eq.~\ref{eq:analytic_real}. Solid curves in (B) and (D) show Eq.~\ref{eq:analytic_scaled}.
        }
    \label{fig:Power_velocity}
\end{figure*}

Having determined that continuous sampling and continuous ratcheting maximize the extracted power, we explored the role of bead mass in experiments using nominal bead diameters of 0.5, 1.5, 3, and 5 \textmu m. For each trap strength $\kappa$ (set by the trapping laser power) and for each trapped bead (whose size varies slightly from the nominal size listed by the manufacturer), we determine the value of feedback gain $\alpha$ that makes $P_{\mrm{trap}} \approx 0$.  The gray markers in Fig.~\ref{fig:Power_velocity}\ti{D} show that the trap power can be kept small, even though the required value of $\alpha$ is different for each value of $\dgr$.

We then measured the extracted velocity (Fig.~\ref{fig:Power_velocity}\ti{A}) and power (Fig.~\ref{fig:Power_velocity}\ti{C}) for the three nominal bead sizes. For a fixed bead size, the power and velocity increase monotonically with trap stiffness, as increasing the trap stiffness reduces the mean first-passage time. 
For fixed trap stiffness and bead density, the velocity decreases with bead size. By contrast, the power is maximized at an optimal intermediate bead size.  Scaling the length, time, and energy by the trap standard deviation $\sigma$, trap relaxation time $\trelo$, and $\kT$, respectively, collapses the data onto single scaled power and velocity curves as a function of the scaled effective mass $\dgr$, Eq.~\ref{eq:analytic_scaled} (Fig.~\ref{fig:Power_velocity}\ti{B} and \ti{D}).  The power is maximized at $P^{*} \approx 0.295\ \kT/\trelo$ for $\dgr\approx 0.845$ and the velocity at $v^{*} \approx 0.8\ \sigma/\trelo$ for $\dgr\to 0$~\cite{park2016}.  The maximum in extracted power at finite $\delta_g$ results from a competition between two effects: the potential energy of a raised object increases with mass, but so does the time to fluctuate beyond a threshold. 

Finally, we explored the influence of gravity on directed motion. Our analysis suggests that gravitational effects should be quantified by the scaled effective mass, $\dgr$. To test this idea, we compared the measured directed velocity achieved for horizontal motion with that achieved for vertical motion (\SI, section \ref{App:G}). Horizontal velocities are consistent with predictions based on Eq.~\ref{eq:analytic_real_velocity} or, equivalently Eq.~\ref{eq:analytic_scaled} for $\dgr \to 0$.  Thus, when the bead is sufficiently light or small (1.5~\textmu m in this case), gravity becomes irrelevant: particle speed is independent of direction (\SI, Fig.~\ref{fig:Hor_vs_ver}). For heavier beads and smaller spring constants, the motion is slower in the vertical direction.

\section*{Discussion}

We have designed a simple information-fueled engine that can convert the heat of a surrounding bath into directed motion and hence store gravitational potential energy. A systematic study of conditions that optimize the performance limits of the engine shows that continuous measurements and continuous ratcheting are best. Fortunately, the analysis of the continuous-feedback limit is simpler than that for the corresponding discrete-time dynamics and can draw on well-known results from the analytic theory of mean first-passage times. From the optimization, we find simple expressions for extracted power and velocity establishing that the performance limits of the engine are set by material parameters such as the stiffness of the spring created by the optical tweezers.

Figure~\ref{fig:Power_velocity}\ti{A,C} and Eq.~\ref{eq:analytic_real} show that smaller beads maximize directed motion, but larger beads maximize power extraction. That varying goals call for varying design principles is familiar in macroscopic applications. For example, the diesel engines used in trucks are optimized for power, whereas the turbocharged engines used in race cars are optimized for speed. More generally, systematic connections between material parameters and performance limits are common features of motors. Indeed, motors ranging from proteins to jet engines follow scaling laws whose form is determined by the failure modes of the materials used in the motor construction~\cite{marden2002,hess2018}.

By following optimal design principles, we have markedly improved performance relative to previous efforts, which focused instead on information-processing costs and the associated ``information-to-work'' efficiency of the engine~\cite{ribezzi2019,admon2018,paneru2018prl,toyabe2010}. The maximum extracted power is $10^4$ times higher than that reported in Ref.~\cite{admon2018}, although comparable laser powers are used. Most of the improvement in extracted power is achieved through the trap design. In the present case, power is applied where needed, via a single trap; an array of traps was used in \cite{admon2018}. Our design may also be compared with \cite{paneru2018prl}, which uses a single trap, as here, but does not store work. The power levels achieved here exceed those in \cite{paneru2018pre} by an order of magnitude. 
The improvement relative to \cite{paneru2018pre}  arises from careful optimization of parameters (bead size, $\xT$, etc.). Similarly, we increase the directed velocity by a factor of $30$ compared to \cite{lee2018scirep} by choosing a smaller bead. 

For our setup, the ``best'' values achieved for power and velocity are 1066 $\kT$/s and 190 \textmu m/s, respectively. These values are significant: They are roughly ten times faster than \ti{E.~coli} and are comparable to the speeds of faster motile bacteria such as those found in marine environments (who need to outswim their algae prey)~\cite{barbara03} and are also comparable to the power used to drive molecular motors such as kinesin~\cite{ariga2018}. 

For setups similar to the one used here, the laser power can in principle be increased significantly, which would increase the trap constant $\kappa$; however, in many applications, heating will limit the power that can be applied. Another route to increasing performance is to optimize the response properties of the trapped particle. Here, we limited our particle choice to dielectric spheres; more sophisticated core-shell particle designs can reduce beam reflection and scattering forces, thereby increasing the trap stiffness at fixed laser power by a factor of approximately ten~\cite{jannasch2012}.

In our experiments, the optical-tweezer setup imposed a harmonic potential. Could more power or higher velocities be possible using a different potential shape? We numerically studied a potential with controllable asymmetry and found no improvement, given a fixed maximum stiffness.  Additionally, we can show that, for symmetric traps, the harmonic shape is optimal (\SI, section \ref{App:K}).

Beyond technological limits set by the stiffness of the material used to build the motor, the dynamical model used in our optimization can break down.  Naively, decreasing dynamical time scales (e.g., by increasing the trap stiffness $\kappa$) always improves information-engine performance.  However, our analysis assumes Eq.~\ref{eq:EQM_full}, which describes a simple overdamped Langevin model with instantaneous damping and is characterized by the relaxation time $\trelo$. For the range of $\kappa$ and bead sizes that we explore, this assumption holds; however, as $\kappa$ increases, the time scale $\trelo$ of the trap dynamics decreases.

If short enough, other dynamical time scales, linked to inertial and memory effects in the surrounding fluid, can act to filter high-frequency fluctuations, thereby limiting the ratcheting achievable through feedback that is based solely on the most recent measurement. To capture inertial effects, the term $m\ddot{x}$ should be included in Eq.~\ref{eq:EQM_full}, which introduces the velocity relaxation time scale $\tau_{v} = m/\gamma$.  
To capture memory effects, the viscous friction term $\gamma \dot{x}$ generalizes to a convolution with a kernel that captures the effects of fluid rearrangements in response to bead motion.  
This introduces a time scale $\tau_{\mrm{f}} = r^{2}/\nu$, the time it takes the fluid to diffuse one particle radius $r$, where $\nu$ is the kinematic viscosity.
The combined effects of inertia and hydrodynamic memory are captured by the Basset-Boussinesq-Oseen equation~\cite{seyler2019}.  

We have made informal numerical studies of these two effects.  
On the one hand, we find that our proposed feedback algorithm (Eq.~\ref{eq:feedback-alg}) leads to worse performance than implied by estimates based on the overdamped limit. The performance begins to degrade at trap dynamics time scales $\approx 3$ \textmu s, obtained by equating the overdamped relaxation time to the fluid memory time scale $\trelo = \tau_{\mrm{f}}$.
This regime is achieved by the 5 \textmu m bead at a trap stiffness of $\kappa \approx 6000$ pN/\textmu m, which is about an order of magnitude greater than our current setup is capable of; however, deviations are empirically already seen for $\kappa \approx 200$ pN/\textmu m (\SI, Figure S6).
Nevertheless, ``naively'' extrapolating the overdamped theory to this time scale implies work extraction of $\approx 10^{4}\ \kT/\mrm{s}$ (for a 5 \textmu m bead) and speeds of $\approx 3000$ \textmu m/s (for a 0.5 \textmu m bead). See \SI, section \ref{App:L}.

On the other hand, these new physical effects are characterized by new dynamical variables that can be used to further optimize the feedback algorithm.  
When inertial effects are important, measuring the velocity $\dot{x}$ can improve feedback; 
likewise, when the hydrodynamic memory kernel is important, the history of positions can help.  In principle, one could modify the feedback rule to incorporate the recent history rather than just the most recent measurement.  Operating an information engine in a gas of reduced pressure~\cite{tebbenjohanns20} would make these scales more accessible experimentally, and it would be interesting to explore whether improved algorithms can capture some of the performance that would otherwise be lost in these regimes.  

Although our focus here has been on information engines supplied with low-noise measurements, it would be interesting to study performance optimization when information costs are considered. 
As noted in the introduction, previous studies have measured information-to-work conversion efficiencies~\cite{ribezzi2019,admon2018,paneru2018pre,paneru20}; however, there has been no systematic study of optimal algorithms -- just particular case studies.

Finally, our information-engine design exploits only the ``up'' fluctuations.  In Szilard's original proposal, the ability to change the connection between mass and partition as a function of the measurement outcome (the side on which the particle is found) allowed exploitation of \ti{all} measurement outcomes. But in our design, ``down'' fluctuations lead to no feedback response. The information gathered in measuring those fluctuations cannot be exploited, reflecting a structural limitation of the engine~\cite{still2020}. A design that could convert and store energy from all measurements would further enhance information-engine performance.

\section*{Materials and Methods}
\subsection*{Experimental setup}
The experiments were performed using an optical-tweezer setup that can rapidly shift the beam position under feedback control~\cite{kumar2018nanoscale,albay18}.  For setup details, see \ti{Appendix}, section \ref{App:A}.

\subsection*{Data Analysis}
To estimate the power and velocity from empirical data, we record trajectories over a fixed distance of 340 nm, a range set by the quadrant photodiode sensor, which records beam deflections due to bead movement.  Every time the bead reaches the upper bound, it is returned to the lower bound, and the ratchet protocol is repeated. Each 340-nm trajectory contains $N_\mrm{ratch} \approx 80$ ratchet events when the threshold $\xT=0$. Typically, the first relaxation time $\tau_\mrm{r}$ of the trajectory is not included when estimating power and velocity, to allow the system to reach steady state. The total displacement and time for each trajectory $j$ is recorded, and then the procedure is repeated $N_\mrm{traj} \approx  100$ times.  The velocity and power are calculated from the average over the $N_\mrm{traj}$ trajectories as $\overline{v} = \sum \left(x_n-x_{n-1}\right) / \ts$ and $\overline{P} = \dgr \overline{v}$, where the sum is over time steps within a trajectory \ti{and} over the $N_\mrm{traj}$ multiple trials.  Because the total number of ratcheting events is large ($N_\textrm{tot} = N_\mrm{ratch}\times N_\mrm{traj} \approx 8 \times 10^3$) and each first-passage time $\tau_\mrm{FP}$ is an independent random variable, we can aggregate the first-passage times from all $N_\textrm{tot}$ events. As $N_\textrm{tot}\gg 1$, 
the law of large numbers can be used to estimate the mean velocity and power. The approximation becomes exact when $N_\mrm{ratch},N_\mrm{traj} \to \infty$ (\ti{Appendix}, section \ref{App:I}).  
\subsection*{Sample preparation} Four sizes of silica bead were used, with nominal diameters specified by the manufacturer of 1.49 $\pm$ 0.22 \textmu m (Bangs Labs), 0.50 $\pm$ 0.05 \textmu m, 3.00 $\pm$ 0.25 \textmu m, and 5.00 $\pm$ 0.35 \textmu m (Sigma-Aldrich). The properties (diffusion constant and force constant) associated with each bead were measured individually before each set of experiments done with the particular bead. The sphere solution from the manufacturer was diluted using deionized water. The sample chamber was prepared from a glass slide and a coverslip, which were separated by 100-\textmu m spacer wires and sealed by nail polish. For the 0.5-\textmu m-bead experiment, the sample chamber consisted of two coverslips, separated by 50-\textmu m spacer wires.

\begin{acknowledgments}
We thank Avinash Kumar and Luis Reinalter (SFU Physics) for contributions to the experimental setup, and Susanne Still (U.~Hawaii) for fruitful discussions. This research was supported by grant number FQXi-IAF19-02 from the Foundational Questions Institute Fund, a donor-advised fund of the Silicon Valley Community Foundation.  Additional support was from Natural Sciences and Engineering Research Council of Canada (NSERC) Discovery Grants (D.A.S and J.B.), a Tier-II Canada Research Chair (D.A.S), an NSERC Undergraduate Summer Research Award (J.N.E.L), a BC Graduate Scholarship (J.N.E.L), and an NSERC Canadian Graduate Scholarship - Masters (J.N.E.L). Computational support was provided by WestGrid and Compute Canada Calcul Canada.
\end{acknowledgments}

\section*{Appendix}
\appendix
\beginsupplement

\section{Experimental apparatus}
\label{App:A}
\begin{figure}[ht]
    \centering
    \includegraphics[width=0.9\linewidth]{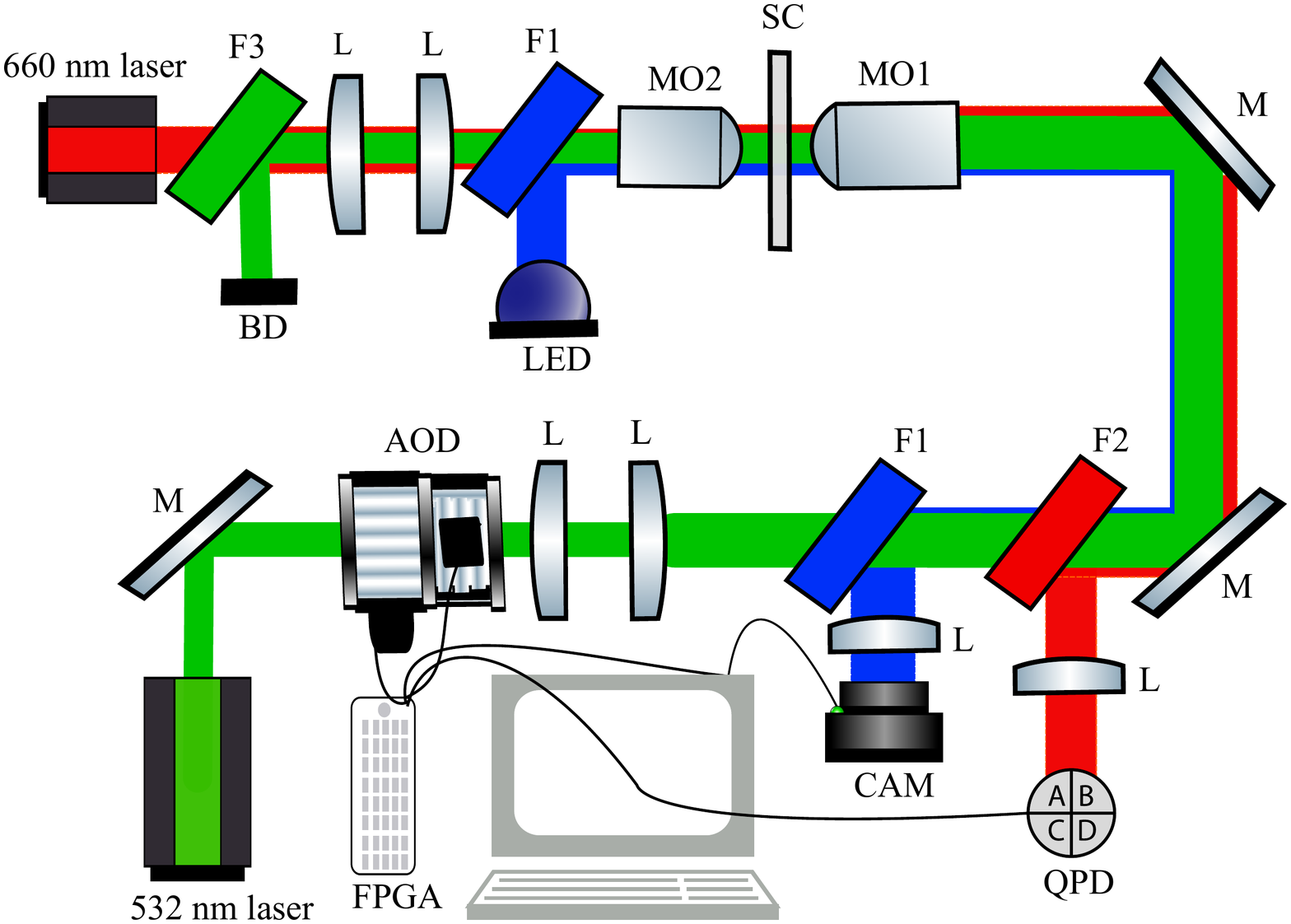}
    \caption{Schematic diagram of the experimental apparatus. M = mirror, AOD = acousto-optic deflectoor, L = lens, F1 = blue dichroic filter, CAM = camera, F2 = red dichroic filter, MO1 = trapping microscope objective, MO2 = detection microscope objective, SC = sample chamber, F3 = green dichroic filter.}
    \label{fig:setup}
\end{figure}

Figure~\ref{fig:setup} shows the schematic of the experimental apparatus, whose basic configuration has been described previously~\cite{kumar2018nanoscale}. The optical trap is made using a 532-nm, green, ring-cavity laser (H\"UBNER Photonics, Cobolt Samba, 1.5 W, 532 nm). 
Two acousto-optic deflectors (AODs, DTSXY-250-532, AA Opto Electronic) can deflect the beam in two directions, to shift the trap's position along both $x$- and $y$-axes. The laser propagates along the $z$-axis; the $x$-axis is parallel to the gravitational force; and the $y$-axis is perpendicular to both gravity and the laser beam. 
The AOD's plane is imaged onto the back aperture of the trapping objective (MO1, water-immersion 60x Olympus objective 1.2 NA) using a two-lens relay system. 
The two lenses also magnify the green laser to overfill the back aperture of the trapping objective. 
The green laser is focused inside the sample chamber (SC), which is filled with a solution of micron-sized beads. Objective MO1 creates the optical trap. 
The AODs also change the power of the trapping laser, to control trap stiffness.

A 660-nm fiber pig-tailed red laser (Thorlabs, LP660-SF20, 20 mW, 660 nm) was used for detection. 
The diameter of the laser beam was kept smaller than the back aperture of the second objective (MO2, 40x Nikon). The reduced beam size leads to a larger focus spot size in the trapping plane, which increases the linear range of the position measurements. 
The detection plane was adjusted using the relay lens (L) system between the red laser and MO2. 
The detection beam is collected by MO1 and focused on the quadrant photo-diode (QPD, First Sensor, QP50-6-18u-SD2). 
The detection laser is reflected on the QPD using a red dichroic filter (F2). 
This signal is used to detect the position of the trapped bead. 

A blue LED was introduced in the laser's path using a high-pass filter (F1), in order to visualize the trapped bead. 
The collected blue light is focused onto a USB-3 digital camera (BASLER, ace acA800-510\textmu m) to image the trapped bead. 
Finally, the green laser that is collected by MO2 is reflected by the green dichroic filter (F3) and then directed to a beam dump (BD).

The voltage from the QPD is sent to the Analog /Digital input of the Field Programmable Gate Arrays (FPGA,  National Instruments, NI PCIe-7857), which converts it to a discretized position signal.
The voltage-to-position conversion factor is obtained by calibrating the QPD-AOD-CAM system (details can be found in \cite{kumar2018spie}). 
The FPGA then makes the feedback decision for the trap reset based on the feedback rule given in the main text. The feedback loop time was 20 \textmu s. In addition to the loop time, the \textit{feedback latency} (time between acquisition of a measurement and subsequent shift of the trap) is also important~\cite{jun2012}. Here, it is always set equal to the sampling time. The feedback latency and scaled effective mass $\dgr$ both affect the value of feedback gain $\alpha$ needed to meet the zero-work condition. We typically find $\alpha$ in the range 1.3--1.9. 

Our theoretical model implies that the directed velocity of the information engine can be increased by reducing the bead size. To test this prediction, we performed experiments with 0.5-\textmu m beads, for which the apparatus had to be modified. A 1.2 NA water-immersion objective was used to focus a higher-power detection laser (H\"UBNER Photonics, Cobolt 06-MLD, 50 mW) to increase the signal from the bead. The maximum velocity and power were obtained for a trap stiffness $\kappa \approx 200$~pN/\textmu m and a (bead-dependent) dynamical time scale as short as $\trelo \approx 23 $ \textmu s, achieved with the maximum trapping laser power of 0.4 W at the trapping plane.

\section{Zero-work condition}
\label{App:B}
The work done by the trap is defined, in scaled units, as
\begin{align}
    W_{n+1} = \frac{1}{2}\left[\pr{x_{n+1}-\lambda_{n+1}}^{2}-\pr{x_{n+1}-\lambda_{n}}^{2}\right].
\end{align}
As the trap position is updated after a delay of one time step, the trap update for time step $n+1$ is based on the bead and trap position at time step $n$. 
Under the assumption that the bead is stationary during the trap update, the trap update that sets the work to zero is given by
\begin{align*}
     0 &= \frac{1}{2}\left[\pr{x_{n}-\lambda_{n+1}}^{2}-\pr{x_{n}-\lambda_{n}}^{2}\right] \\
     \pr{x_{n}-\lambda_{n+1}} &= \pm \pr{x_{n}- \lambda_{n}}\\
     \lambda_{n+1} &= \lambda_n+ 2\pr{x_{n}- \lambda_{n}},
     \label{eqn:feedback}
\end{align*}

which corresponds to a feedback gain $\alpha = 2$.  Figure~2 shows that, experimentally, the zero-work condition is observed for $\alpha \approx 1.5$. This small discrepancy can be explained by time delays in the experiment: There is a one-sample lag ($t_\mrm{s} = 20$ \textmu s) in the experiment that is not accounted for in the simple model, which reduces the value of $\alpha$ required for zero work.

\section{Mean first-passage time}
\label{App:C}
\label{ssec:MFP_fullcalculation}

\begin{figure}[ht]
    \centering
    \includegraphics[width=6cm]{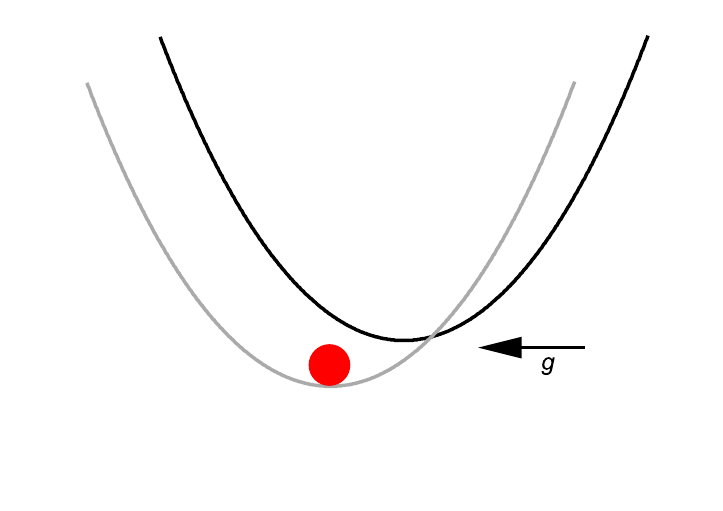}
    \caption{
        Bead at the minimum-energy state of
        the total potential (gray curve) due to the trap and gravity. The black plot is the trap potential. 
    }
    \label{fig:Fullpotential}
\end{figure}

The particle's position propagator $p(x,t|x_0,0)$ obeys the (forward) Fokker-Planck equation: 
\begin{align}
    \pdv{t} p(x,t|x_0,0) &= \nonumber\pdv{x}\left(p(x,t|x_0,0)\pdv{x} V(x)\right)\\
    &\quad+ \pdv[2]{x}p(x,t|x_0,0)\ ,
\end{align}
for total potential
$V(x) \equiv V_\mathrm{t}(x)  + \dgr\,x$ consisting of the sum of the trap potential $V_\mathrm{t}(x) = \frac{1}{2}(x-\lambda)^2$ and the gravitational potential, as shown in Fig.~\ref{fig:Fullpotential}.

In general, the mean first-passage time $\tau_{\mrm{MFP}}(x;b)$ gives the mean time until a particle started at $x<b$ first reaches $b$. It is given by~\cite{haenggi1990}:
\begin{align}
    \tau_{\mrm{MFP}}(x;b) = \int_{x}^{b}\dd{y}\ \e^{V(y)}\int_{-\infty}^{y} \dd{z}\ \e^{-V(z)}\ . \label{eq:FPT}
\end{align}

In our case, we are interested in the MFPT $\tau_{\mrm{MFP}}(X_\mrm{T})$ from $\lambda-X_\mrm{T}$ to $\lambda+X_\mrm{T}$.
Substituting $x=\lambda-X_\mrm{T}$ and $b=\lambda + X_\mrm{T}$
into
\eqref{eq:FPT} 
gives
\begin{align}
    \tau_{\mrm{MFP}}(X_\mrm{T}) 
    &=\nonumber \int_{\lambda-X_\mrm{T}}^{\lambda+X_\mrm{T}} \dd y \,\exp\left\{V_\mathrm{t}(y-\lambda)+\delta_\mrm{g}y\right\} \\
    &\quad \times \int_{-\infty}^{y} \dd z\, \exp\left\{-V_\mathrm{t}(z-\lambda)-\delta_\mrm{g}z\right\}\\
    &=\nonumber \int_{\lambda-X_\mrm{T}}^{\lambda+X_\mrm{T}} \dd y \ \exp\left\{V_\mathrm{t}(y-\lambda)+\delta_\mrm{g}(y+\lambda)\right\}\\
    &\quad \times \int_{-\infty}^{y} \dd z\, \exp\left\{-V_\mathrm{t}(z-\lambda)-\delta_\mrm{g}(z+\lambda)\right\}\\
    &=\nonumber \int_{-X_\mrm{T}}^{X_\mrm{T}} \dd x' \ \exp\left\{V_\mathrm{t}\left(x'\right)+\delta_\mrm{g}x'\right\}\\
    &\quad \times\int_{-\infty}^{x'} \dd x''\, \exp\left\{-V_\mathrm{t}\left(x''\right)-\delta_\mrm{g}x''\right\}\,, \label{eq:generalTrappingMFPT}
\end{align}
where in the last line we substituted $x'' \equiv z-\lambda$ and $x' \equiv y-\lambda$.

\section{Low sampling-frequency limit}
\label{App:D}
When the (scaled) sampling frequency $f_\mrm{s} = \tau_\mrm{r}/t_\mrm{s}$ is small, the position distribution equilibrates between successive feedback steps. It is then given by the Boltzmann distribution (in scaled units)
\begin{align}
    p_\mrm{eq}(x;\lambda) = \frac{1}{\sqrt{2\pi}} \exp\left\{-\tfrac{1}{2}(x-\lambda-\delta_\mrm{g})^2\right\}\,.
\end{align}

In that limit, the average work $W_\mrm{eq}$ extracted per feedback step is determined from the feedback rules given in Eq.~4 in the main text: 
\begin{align}
    W_\mrm{eq} &=\nonumber \delta_\mrm{g}\, \Big[ \underbrace{\phantom{\int} 0 \phantom{\int}}_{\text{positions left of threshold}}\Big]\\
    &\quad +\Big[ \underbrace{\int_{\lambda}^\infty \mrm{d}x\, 2(x-\lambda)p_\mrm{eq}(x;\lambda)}_{\text{positions right of threshold}} \Big]\\
    &= \sqrt{\frac{2}{\pi}}\,\delta_\mrm{g} \int_{\delta_\mrm{g}}^\infty \mathrm{d}x'\,(x'-\delta_\mrm{g})\, e^{-(x')^2/2}\\
    &= \delta_\mrm{g} \left\{ \sqrt{\frac{2}{\pi}} \, e^{-\delta_\mrm{g}^2/2} + \delta_\mrm{g}\left[ \erf\left(\frac{\delta_\mrm{g}}{\sqrt{2}}\right) -1 \right] \right\}\,,
\end{align} 
where in the second line we substituted $x' \equiv x-\lambda-\delta_\mrm{g}$.

Therefore, in the limit of low sampling frequency, the power in scaled units is
\begin{align}
    P &= f_\mrm{s} \, W_\mrm{eq}\\
    &= f_\mrm{s} \,\left\{ \sqrt{\frac{2}{\pi}}\,\delta_\mrm{g} \, e^{-\delta_\mrm{g}^2/2} + \delta_\mrm{g}^2\left[ \erf\left(\frac{\delta_\mrm{g}}{\sqrt{2}}\right) -1 \right] \right\}\,,
\end{align}
or $P \approx 0.19 f_\mrm{s}$ for $\dgr = 0.8$, the value used in Fig.~3\textit{A}.

\section{Arbitrary sampling frequency}
\label{App:E}
To calculate velocity and power for arbitrary sampling frequencies, we first derive a self-consistency equation for the steady-state position distribution as a function of sampling frequency, which we evaluate numerically. 

\begin{figure}[ht]
    \centering
    \includegraphics[clip, width=0.8\linewidth]{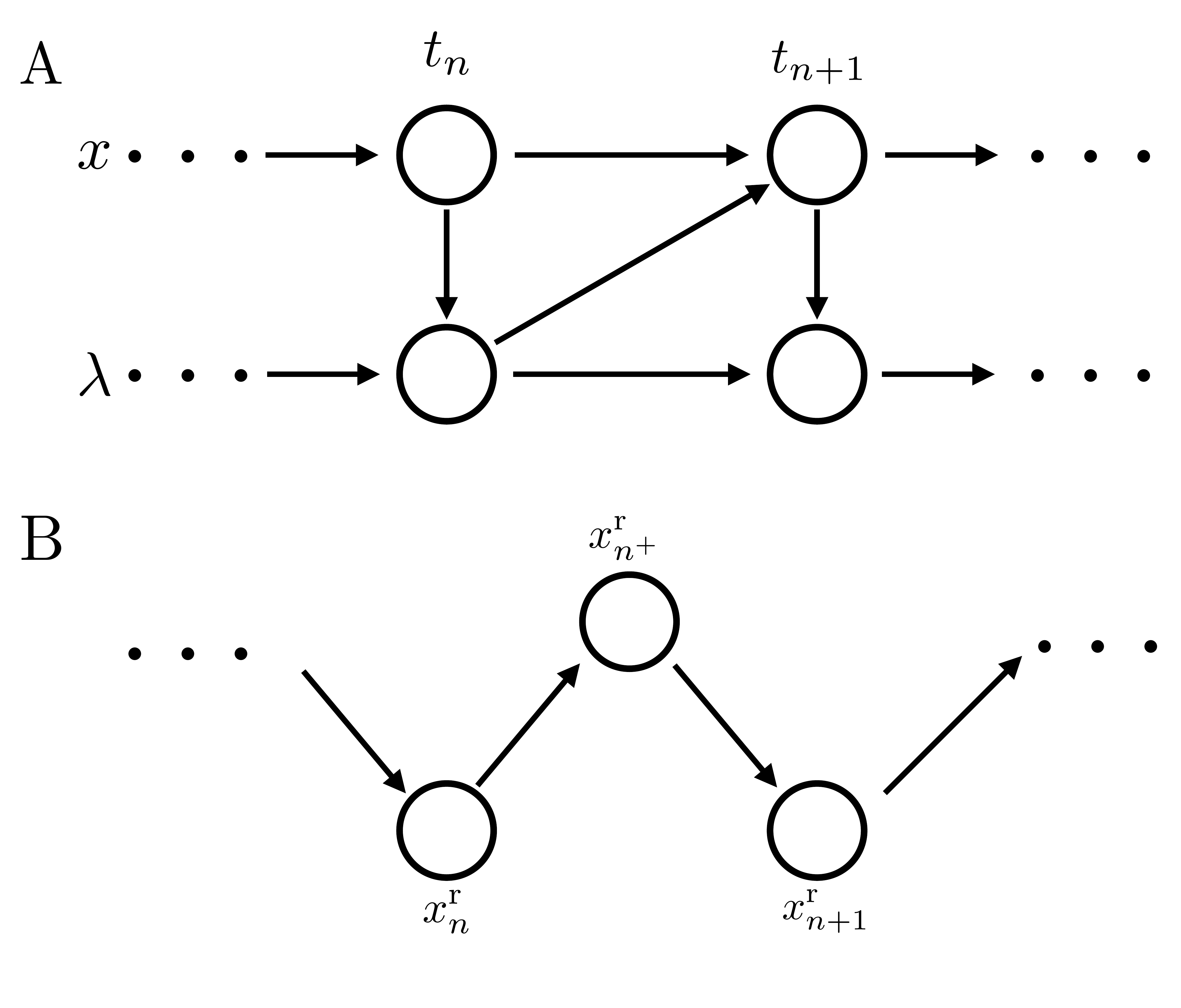}
    \caption{ 
        Causal structure of bead ($x$) and trap ($\lambda$) dynamics. 
        A) Structure of dynamics for true position $x$ and trap center $\lambda$. 
        B) Structure of dynamics for the relative coordinates $\xnr$ and $\xnpr$ defined in \eqref{eq:trafo-var}.
        }
    \label{fig:causal_struct}
\end{figure}

We begin by noting the causal structure given by Fig.~\ref{fig:causal_struct}A. From this causal structure we derive the following decomposition for the joint transition probability to go from bead position $x_n$ and trap center $\lambda_n$ at sampling time $t_n$ to $x_{n+1}$ and $\lambda_{n+1}$ at sampling time $t_{n+1}$:
\begin{align}
    p\pr{x_{n+1},\lambda_{n+1}|x_{n},\lambda_{n}} &=\nonumber p_{x}\pr{x_{n+1}|x_{n},\lambda_{n}}\\
    &\quad \times p_{\lambda}\pr{\lambda_{n+1}|x_{n+1},\lambda_{n}} \, .
\end{align}
The bead-position propagator $p_{x}\pr{x_{n+1}|x_{n},\lambda_{n}}$, in scaled units, 
is given by the infinitesimal generator for an Ornstein-Uhlenbeck process~\cite[Section~5.3]{Risken1996},
\begin{align}
    p_{x}\pr{x_{n+1}|x_{n},\lambda_{n}} &=\nonumber \mcal{N}(x_{n+1}; x_{n}\e^{-\ts} \\
    &\qquad +\left(1-\e^{-\ts}\right)(\lambda_{n}-\dgr),
    1-\e^{-2\ts}) \ , 
\end{align}

where $\mcal{N}(x;\mu,\sigma^{2})$ denotes a normal (Gaussian) distribution over $x$ with mean $\mu$ and variance $\sigma^{2}$. 
The propagator $p_{\lambda}\pr{\lambda_{n+1}|x_{n+1},\lambda_{n}}$ for the trap center is given by
\begin{align}
    p_{\lambda}\pr{\lambda_{n+1}|x_{n+1},\lambda_{n}} &=
    \nonumber \delta\pr{\lambda_{n+1}-\lambda_{n}}\Theta[-(x_{n+1}-\lambda_{n})]\\
    &\nonumber\quad +\delta[\lambda_{n+1}-(2x_{n+1}-\lambda_{n})]\,\\
    &\qquad\times\Theta\pr{x_{n+1}-\lambda_{n}} \ .
\end{align}
Here, $\delta(\cdot)$ denotes the Dirac-delta function, and $\Theta(\cdot)$ denotes the Heaviside function.
The first term on the right-hand side reflects fluctuations that are not sufficient to trigger a ratcheting, 
while the second term  corresponds to ratcheting events.

\begin{figure*}[ht]
    \centering
    \includegraphics[width=\linewidth]{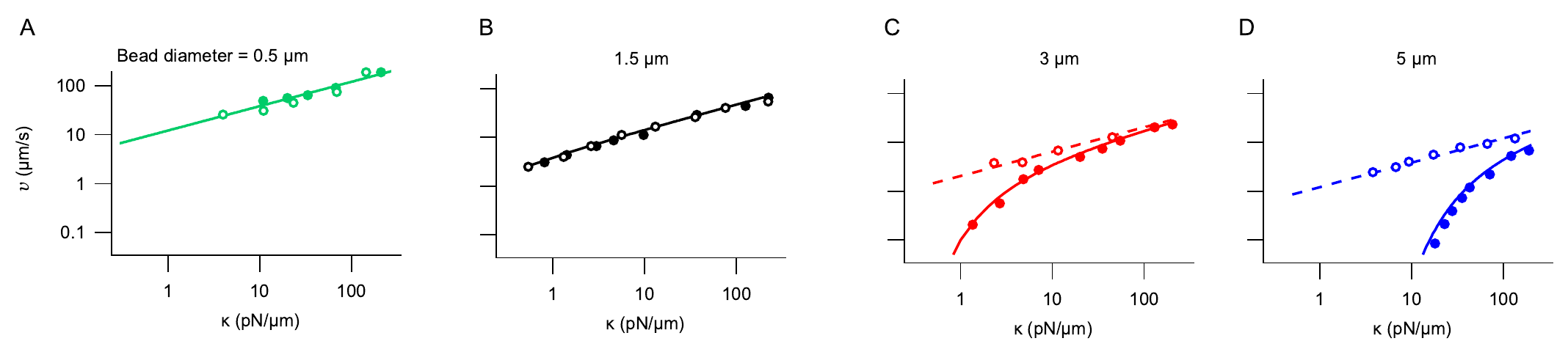}
    \caption{Horizontal and vertical velocity (hollow and solid markers, respectively) as a function of trap stiffness $\kappa$ for different bead diameters. 
        The solid lines correspond to \eqref{eq:vel_real} and reflect the influence of the gravitational force on upward fluctuations.  The dashed lines are calculated from \eqref{eq:vel_hor}. 
        }
    \label{fig:Hor_vs_ver}
\end{figure*}

A change of variables to relative coordinates
\begin{align} \label{eq:trafo-var}
    \xnpr \equiv x_{n+1}-\lambda_{n} \qquad\text{and} \qquad \xnr \equiv x_{n}-\lambda_{n}
\end{align}
yields
\begin{subequations}
    \begin{align}
        p\pr{\xnpr|x_{n},\lambda_{n}} 
        &=\nonumber \int\dd{x_{n+1}} \, \delta[x_{n+1}-(\xnpr+\lambda_{n})]\\
        &\qquad\times p_{x}\pr{x_{n+1}|x_{n},\lambda_{n}}\\
        &=\nonumber \mcal{N}(\xnpr; \ \xnr\e^{-\ts}\\
        &\qquad-\left(1-\e^{-\ts}\right)\dgr, 1-\e^{-2\ts})\\
        &\equiv p_{1}\pr{\xnpr|\xnr} \ .
    \end{align}
\end{subequations}
Similarly, using $\xnnr \equiv x_{n+1}-\lambda_{n+1}$, we obtain
\begin{subequations}
    \begin{align}
        p\pr{\xnnr|x_{n+1},\lambda_{n}} 
        &= \nonumber \int\dd{\lambda_{n+1}} \, \delta[\xnnr-(x_{n+1}-\lambda_{n+1})] \, \\ 
        &\quad \times p_{\lambda}\pr{\lambda_{n+1}|x_{n+1},\lambda_{n}}\\
        &=\nonumber \delta\pr{\xnnr - \xnpr}\Theta\pr{-\xnpr}\\
        &\quad + \delta\pr{\xnnr+\xnpr}\Theta\pr{\xnpr}\\
        &\equiv p_{2}\pr{\xnnr|\xnpr} \ .
    \end{align}
\end{subequations}
Importantly, this change of variables simplifies the causal structure, which is shown in Fig.~\ref{fig:causal_struct}B.

The steady-state solutions $\pi_+\pr{\xnpr}$ and $\pi\pr{\xnr}$ for these two variables $\xnpr$ and $\xnr$, respectively, are given by the self-consistent integral equations
\begin{align}
    \pi_+(\xnpr) &= \int\dd{u}\underbrace{\sr{\int\dd{v} p_{1}\pr{\xnpr|v}p_{2}\pr{v|u}}}_{\equiv T\pr{\xnpr|u}}\pi_+(u).\label{eq:pir_eq}\\
    \pi(\xnr) &= \int\dd{v}\underbrace{\sr{\int\dd{u} p_{2}\pr{\xnr|u}p_{1}\pr{u|v}}}_{\equiv \tilde{T}(\xnr|v)}\pi(v),\label{eq:pis_eq}
\end{align}
where $u$ and $v$ are dummy variables of integration.

The propagator $T\pr{\xnpr|u}$ is given by
\begin{align}
    T\pr{\xnpr|u} 
    &=\nonumber \Theta(-u)\mcal{N}\left(\xnr;(u+\dgr)\e^{-\ts}-\dgr,1-\e^{-2\ts}\right) \\
    & + \Theta\pr{u}\mcal{N}\left(\xnr;-\dgr-(u-\dgr)\e^{-\ts},1-\e^{-2\ts}\right). \label{eq:xnpr_prop}
\end{align}
Similarly, the propagator $\tilde{T}\pr{\xnr|v}$ is given by
\begin{align}
    \tilde{T}\pr{\xnr|v}&=\nonumber \Theta\pr{-\xnr}\mcal{N}\left(\xnr;\dgr-(v+\dgr)\e^{-\ts}, 1-\e^{-2\ts}\right)\\
    & +\Theta\pr{-\xnr}\mcal{N}\left(\xnr;(v+\dgr)\e^{-\ts}-\dgr,1-\e^{-2\ts}\right) . \label{eq:xnr_prop}
\end{align}

We numerically solve for the steady-state distributions, Eqs.~(\ref{eq:pir_eq}) and (\ref{eq:pis_eq}), by discretizing the propagator~\eqref{eq:xnpr_prop} in $\xnpr$ and $u$ and the propagator~\eqref{eq:xnr_prop} in $\xnr$ and $v$.
The associated eigenvectors of $T\pr{\xnpr|u}$ and $\tilde{T}\pr{\xnr|v}$ with eigenvalue 1 give the distributions $\pi_+\pr{\xnpr}$ and $\pi\pr{\xnr}$, respectively.
Here, we use 2000 uniformly spaced grid points in the domain $\xnr,\xnpr,u,v\in[-20,20]$.   

The steady-state output power $P$ in a given cycle can be recast in terms of the new variables $\xnr$ and $\xnpr$ as
\begin{align}
    P = \ev{\Delta W_{\mrm{g}}}\fs = \dgr\pr{\ev{\xnpr}-\ev{\xnr}}\fs \ ,
\end{align}
where $\ev{\Delta W_{\mrm{g}}}$ denotes the work done per cycle, and the averages are taken over the appropriate steady-state distributions for each of the variables.  Figure~3A of the main text compares numerical results using this approach and the experiment.

\section{Large-stiffness asymptotics for power and velocity}
\label{App:F}
As shown in the main text (Eq.~12a,b), power and velocity are given (in physical units) by
\begin{align}
    P' &=\frac{\, k_{\rm B}T}{\tau_r}\,\sqrt{\frac{2}{\pi}}\,\delta_\mrm{g}  e^{-\delta_{\rm g}^2/2}\,\left[ 1 + \erf\left( \frac{\delta_{\rm g}}{\sqrt{2}}\right) \right]^{-1} 
    \label{eq:P_real}\\
    v' &= \frac{\sigma} {\tau_r}\,\sqrt{\frac{2}{\pi}}\,  e^{-\delta_{\rm g}^2/2}\,\left[ 1 + \erf\left( \frac{\delta_{\rm g}}{\sqrt{2}}\right) \right]^{-1}\,,
    \label{eq:vel_real}
\end{align}
where $\tau_{\rm r} = \gamma/\kappa$, $\delta_\mrm{g} = mg/\kappa\sigma$, and $\sigma = \sqrt{\kT/\kappa}$.

An expansion for large trap stiffness $\kappa$ leads to the following asymptotic relations:
\begin{align}
    P' &\sim \sqrt{\frac{2 k_{\rm B}T}{\pi}} \frac{m g\,\sqrt{\kappa}}{\gamma} \,, \\
    v' &\sim \sqrt{\frac{2 k_{\rm B}T}{\pi}} \frac{\sqrt{\kappa}}{\gamma} \,.
    \label{eq:vel_high_kappa}
\end{align}

\section{Horizontal vs.\ vertical velocity}
\label{App:G}
In the horizontal direction, there is no gravitational effect. The bead fluctuates subject solely to the harmonic potential of the trap. Setting $\delta_\mrm{g} = 0$ in \eqref{eq:vel_real} 
gives the velocity
\begin{align}
    v' & = \sqrt{\frac{2 k_{\rm B}T}{\pi}} \frac{\sqrt{\kappa}}{\gamma}\,.
    \label{eq:vel_hor}
\end{align}
Figure~\ref{fig:Hor_vs_ver} shows the velocity along the horizontal and vertical directions as a function of trap stiffness. For the 1.5-\textmu m bead, the horizontal and vertical velocities are indistinguishable.
The system is in the high-$\kappa$ limit, so the vertical velocity is well approximated by Eq.~\ref{eq:vel_high_kappa}. For larger
beads, the vertical velocity asymptotes to the horizontal velocity as the stiffness increases.

\section{Trap stiffness}
\label{ssec:trap_stiffness}

\begin{figure}[ht]
    \centering
    \includegraphics[width = 0.8\linewidth]{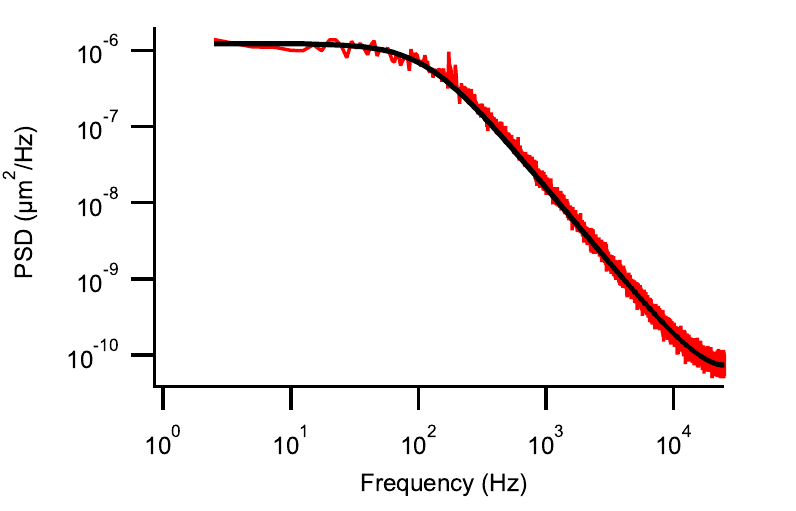}
    \caption{
        Power spectral density for 3 \textmu m bead. 
        The red curve is the experimental data and solid black curve is the fit. 
        The fit parameters are $f_{\rm c} = 113.42 \pm 1.34$ Hz and $D = 0.167 \pm 0.001$ \textmu m$^2$/s. 
        }
    \label{fig:power_spec}
\end{figure}

The trap stiffness $\kappa$ is measured by fitting the power spectrum of the position data to an $aliased$ $Lorentzian$ \cite{soren2004}. The fit parameters, the diffusion constant $D$ and corner frequency $f_{\rm c}$,
are used to evaluate the trap stiffness. The quantities are related by
\begin{align}
    \kappa = \frac{2 \pi f_c\, k_{\mrm{B}}T}{D}.
\end{align}
Figure~\ref{fig:power_spec} shows the power spectrum and the fit.

\section{Optimal bead size that maximizes
power}
\label{App:I}

\begin{figure}[ht]
    \centering
    \includegraphics[width = \linewidth]{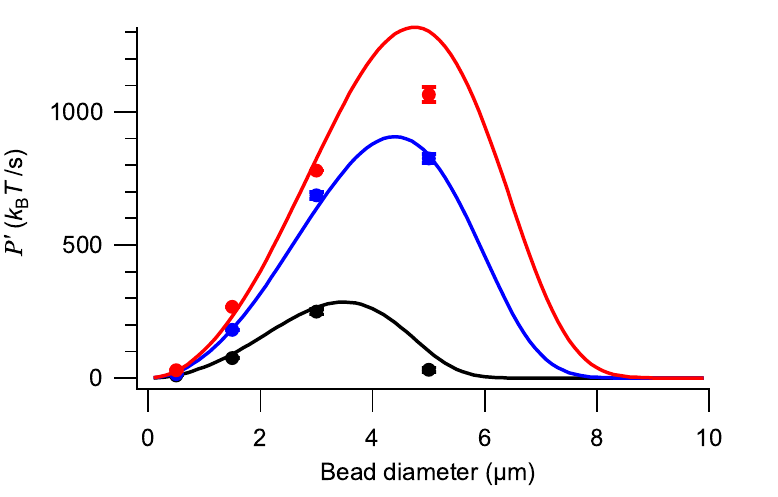}
    \caption{Optimum bead size.
        Power as a function of 
        bead size for trap stiffness $\kappa = 30$ pN/\textmu m (black), $\kappa = 100$ pN/\textmu m (blue) and $\kappa = 200$ pN/\textmu m (red). The circles are experimental data. Curves show Eq.~\ref{eq:P(r)}.
        }
        \label{fig:PvsBeadDiameter}
\end{figure}

Figure~4\textit{C} shows that there is an optimal $\delta_\mrm{g}$ that maximizes the power. 
The optimization can be understood in physical units as well, where we 
change the bead size
subject to fixed trap stiffness and bead density. To study the size dependence, Eq. \ref{eq:P_real} can be written as
\begin{multline}
   P' =\frac{\, \sqrt{\kappa k_{\rm B}T}}{6\eta}\,
    \sqrt{\frac{2}{\pi}}\,
    \frac{4 r^2\Delta\rho g}{3}\,
    \exp\left\{-\frac{1}{2}\left(C\frac{ r^3}{\sqrt{\kappa}}\right)^2\right\}\,\\
    \times \left[ 1 + \erf\left( \frac{1}{\sqrt{2}}\left(C\frac{ r^3}{\sqrt{\kappa}}\right)\right) \right]^{-1} \ ,
    \label{eq:P(r)}    
\end{multline}

where $C = \frac{4\pi \Delta\rho}{3}\frac{g}{\sqrt{k_{\rm B}T}}$, $\Delta \rho$ is the effective density of the bead in water, $\eta$ the viscosity of water, and we have used the relations $\tau_\mrm{r} = \gamma/\kappa$, $\gamma = 6\pi \eta r$ and $\sigma = \sqrt{k_{\rm B}T/\kappa}$. 
Figure~\ref{fig:PvsBeadDiameter} shows the extracted power as a function of bead diameter. 
For $\kappa = 30$ pN/\textmu m, the optimal bead diameter is $\approx$ 3.5 \textmu m; when the trap stiffness is increased, the optimum shifts towards larger beads.

\section{Velocity estimator} 
\label{App:J}
We calculate the velocity by averaging over $N_\mrm{traj}$ trajectories indexed by $j$. Each trajectory has a different duration; consequently, the number $N(j)$ of position measurements varies for each trajectory:
\begin{align} \label{eq:estimator_velocity}
    v &= \frac{\sum_{j=1}^{N_\mrm{traj}} \sum_{n=1}^{N(j)} (x_n - x_{n-1})}{\sum_{j=1}^{N_\mrm{traj}} \sum_{n=1}^{N(j)} t_\mrm{s}} \ ,
\end{align}   
where $t_\mrm{s}$ is the sampling time, and $x_n$ denotes the particle position at sampling time $t_n$.

In the following, we derive Eq.~7 from this estimator in the limit of long trajectories.
For $N_\mrm{traj}\gg 1$, the law of large numbers permits us to write
\begin{align}
    \left\langle v \right\rangle &= \frac{N_\mrm{traj}\,\left\langle \sum_{n=1}^{N} (x_n - x_{n-1}) \right\rangle_{p(N)} }{N_\mrm{traj}\, \left\langle \sum_{n=1}^{N} t_\mrm{s} \right\rangle_{p(N)}} \,,
\end{align}
where we average over the random number $N$ of time samples in a trajectory.

We rewrite this expression in terms of the average over the number $N_{\mrm{ratch}}$ of ratchet events during a trajectory,
\begin{align}
    \left\langle v \right\rangle =  \frac{\left\langle \sum_{m=1}^{N_{\mrm{ratch}}} \Delta x_m \right\rangle_{p(N_{\mrm{ratch}})} }{ \left\langle \sum_{m=1}^{N_{\mrm{ratch}}} \tau_\mrm{FP} \right\rangle_{p(N_{\mrm{ratch}})}}\ ,
\end{align}
where $\Delta x_m$ denotes the bead displacement of ratchet event $m$, and $\tau_\mrm{FP}$ is the time interval between successive ratchet events (a first-passage time).

The system quickly reaches a steady state, which implies that all ratchet events are independent and identically distributed events. Steady state is ensured by neglecting the first relaxation time of each trajectory.
Then, with $N_{\mrm{ratch}} \gg 1$, the central limit theorem implies
\begin{align}
    \left\langle v \right\rangle &= \frac{\left\langle N_{\mrm{ratch}}\left\langle \Delta x \right\rangle \right\rangle_{p(N_{\mrm{ratch}})} }{\left\langle  N_{\mrm{ratch}} \left\langle \tau \right\rangle \right\rangle_{p(N_{\mrm{ratch}})}}\ ,
\end{align}
where the inner averages $\ev{\cdot}$ are now taken over a single ratchet event. 

Now, $\langle \Delta x\rangle = X_\mrm{T} +X_\mrm{R}$ is an exact relation in the fast-sampling limit, which also implies that the average time to ratchet (i.e., reach $X_\mrm{T}$ from $-X_\mrm{R}$) is given by the mean first-passage time:  $\left\langle\tau\right\rangle \to \tau_\mrm{MFP}$. Thus, the velocity relation in Eq.~\ref{eq:estimator_velocity} leads to Eq.~7: 
\begin{align} 
    \left\langle v \right\rangle & = \frac{X_\mrm{T} + X_\mrm{R}}{\tau_\mrm{MFP}} \ .
\end{align}

\section{Non-harmonic and asymmetric potentials}
\label{App:K}
Here, we discuss the influence of the shape of the trapping potential on the maximum extracted power and provide support for the claim made in discussion section of the main text that choosing a potential whose shape differs from a quadratic cannot improve the performance of an information engine. To restrict the discussion to physically relevant settings, we assume that the material used to construct the trap has a maximum stiffness. Otherwise, it would be possible to extract work at infinite rate with an infinitely stiff (\emph{wall-like}) potential.

We first investigate whether asymmetric trapping potentials can increase the power above that achieved by symmetric traps.

\subsection{Numerical exploration of asymmetric trapping potentials}
\begin{figure}
    \centering
    \includegraphics[width = 0.9 \linewidth]{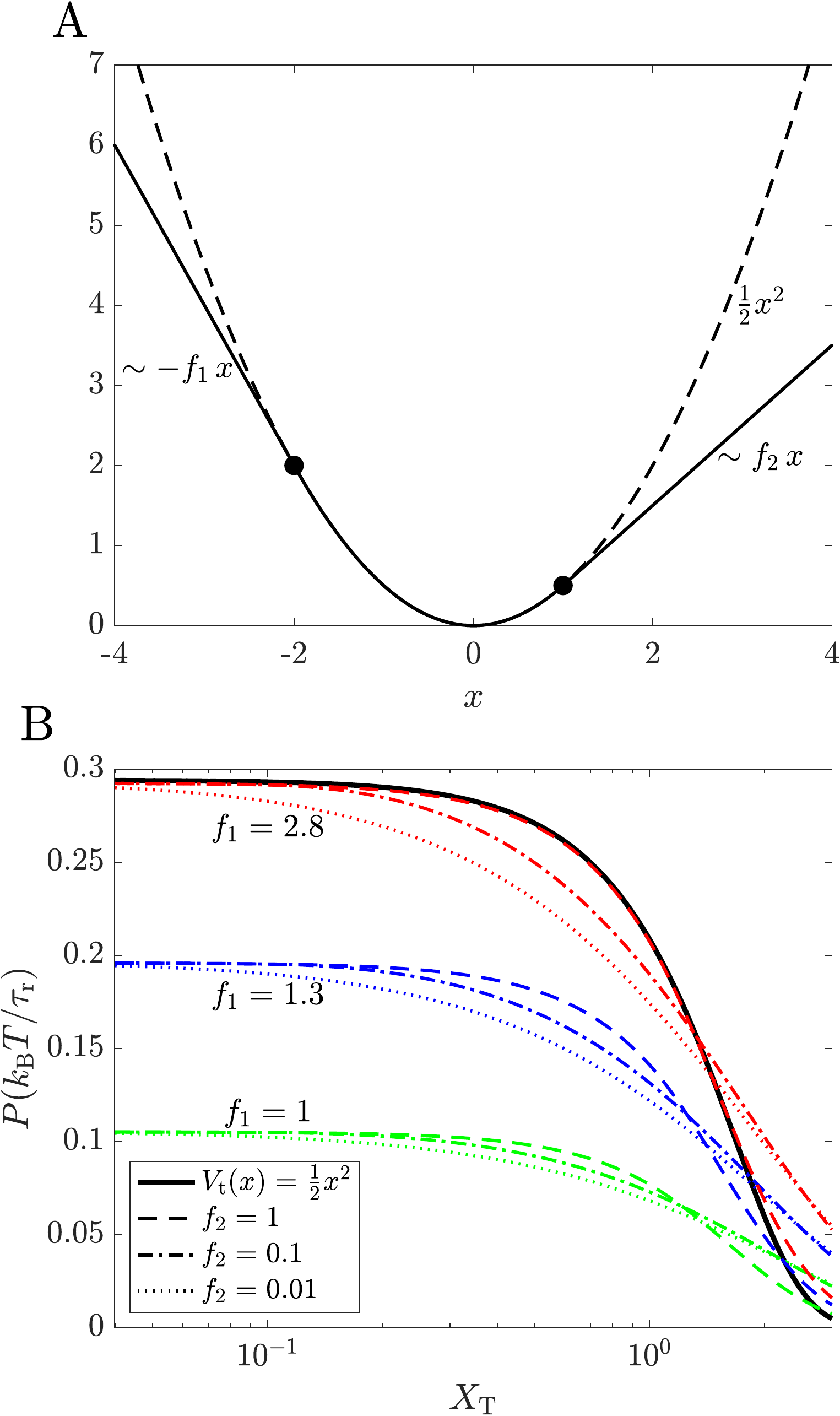}
    \caption{The \emph{quadratic-to-linear} trap. A) Sketch of the potential for 
    maximum slope magnitudes $f_1=2$ for negative $x$ and $f_2=1$ for positive $x$. 
    B) Power as a function of threshold for different maximum slope magnitudes $f_1$ and $f_2$ and $\delta_\mrm{g}=0.8$.}
    \label{fig:quadratic_to_linear}
\end{figure}

To study whether an asymmetric trapping potential can increase the power 
even further, beyond an equivalent symmetric trap, we consider the following \emph{quadratic-to-linear} trap:
\begin{align}
    V_\mathrm{t}(x) &= \begin{cases}
    -f_1 x - \frac{f_1^2}{2}, & x < -f_1\\
    \frac{1}{2}x^2, & -f_1 \leq x < f_2 \\
    f_2 x - \frac{f_2^2}{2}, & f_2 \leq x\,,
    \end{cases}
\end{align}
which describes an asymmetric continuous and differentiable trapping potential that is linear at large displacements
with a quadratic minimum (see Fig.~\ref{fig:quadratic_to_linear}A). Importantly, this potential can never exceed the harmonic trap (which corresponds to $f_1,f_2 \to \infty$), and its curvature is always less than or equal to that of a harmonic trap with potential $\tfrac{1}{2}x^2$.

The condition of zero work done by the trap implies 
\begin{align}
    X_\mathrm{R}(X_\mathrm{T}) := \begin{cases}
    X_\mathrm{T}, & X_\mathrm{T} < f_2\\
    \sqrt{2 f_2 X_\mathrm{T} - f_2^2}, & f_2 \leq X_\mathrm{T} < \frac{f_1^2+f_2^2}{2 f_2} \\[3pt]
    \frac{f_1^2-f_2^2 + 2f_2 X_\mathrm{T}}{2 f_1}, & \frac{f_1^2+f_2^2}{2 f_2} \leq X_\mathrm{T}\,
    \end{cases}
\end{align}
when $f_1 \geq f_2$ and
\begin{align}
    X_\mathrm{R}(X_\mathrm{T}) := \begin{cases}
    X_\mathrm{T}, & X_\mathrm{T} < f_1\\
    \frac{f_1^2+X_\mathrm{T}^2}{2 f_1}, & f_1 \leq X_\mathrm{T} < \frac{f_2^2}{2} \\[3pt]
    \frac{f_1^2-f_2^2+2 f_2 X_\mathrm{T}}{2 f_1}, & \frac{f_2^2}{2} \leq X_\mathrm{T}\,
    \end{cases}
\end{align}
when $f_2 > f_1$.

The mean first-passage time can be computed by numerically evaluating the integral
\begin{align}
    \tau_\mrm{MFP}(X_\mrm{T}) = \int^{X_\mrm{T}}_{-X_\mrm{R}(X_\mrm{T})} \mrm{d} x\, e^{V_\mrm{t}(x) + \delta_\mrm{g} x} \int_{-\infty}^{x} \mrm{d} y \, e^{-V_\mrm{t}(y) - \delta_\mrm{g} y}\,.
\end{align}
Finally, the power as a function of the threshold reads
\begin{align}
    P(X_\mathrm{T}) = \frac{\delta_g \left[X_\mathrm{R}(X_\mrm{T}) + X_\mrm{T}\right]}{\tau_\mrm{MFP}(X_\mrm{T})}\,.
\end{align}

Figure~\ref{fig:quadratic_to_linear}B shows the power as a function of the threshold, for different slopes $f_1$ and $f_2$. We see that an asymmetric trapping potential does not outperform the harmonic potential. Furthermore, maximum power is still achieved at vanishing threshold and is determined by the maximum slope of the left side of the potential, which prevents downward fluctuations. This conclusion is expected since, at vanishing threshold, the mean first-passage time is independent of the right side of the potential.

An intuitive lesson from this example is that, perhaps unsurprisingly, one should design a trap to minimize unproductive downward fluctuations.

\subsection{For symmetric trapping potentials, power is maximized at vanishing threshold} \label{sec:symmetric_thresholds}
Since symmetric trapping potentials do as well as asymmetric ones, we investigate whether a vanishing threshold is always optimal. Here, we consider a generic symmetric trapping potential $V_\mathrm{t}(x) = V_\mathrm{t}(-x) $ with a well-defined minimum $V(0) = 0$ at $x=0$.

The mean first-passage time is given by [see~\eqref{eq:generalTrappingMFPT}]:
\begin{align}
    \tau_\mrm{MFP}(X_\mrm{T}) = \int_{-X_\mrm{T}}^{X_\mrm{T}} \mrm{d} x\, e^{V_\mrm{t}(x) + \delta_\mrm{g} x} \int_{-\infty}^{x} \mrm{d} y \, e^{-V_\mrm{t}(y) - \delta_\mrm{g} y}\,.
\end{align}
We rewrite this expression by splitting the integral first at $x=0$ and then at $y=0$,
\begin{widetext}
\begin{align}
    \tau_\mrm{MFP}(X_\mrm{T})&=\int_{-X_\mrm{T}}^{0} \mrm{d} x\, e^{V_\mrm{t}(x) + \delta_\mrm{g} x} \int_{-\infty}^{x} \mrm{d} y \, e^{-V_\mrm{t}(y) - \delta_\mrm{g} y}
        + \int_{0}^{X_\mrm{T}} \mrm{d} x\, e^{V_\mrm{t}(x) + \delta_\mrm{g} x} \int_{-\infty}^{x} \mrm{d} y \, e^{-V_\mrm{t}(y) - \delta_\mrm{g} y}\\
    &= \int_{-X_\mrm{T}}^{0} \mrm{d} x\, e^{V_\mrm{t}(x) + \delta_\mrm{g} x} \int_{-\infty}^{0} \mrm{d} y \, e^{-V_\mrm{t}(y) - \delta_\mrm{g} y} 
        - \int_{-X_\mrm{T}}^{0} \mrm{d} x\, e^{V_\mrm{t}(x) + \delta_\mrm{g} x} \int_{x}^{0} \mrm{d} y \, e^{-V_\mrm{t}(y) - \delta_\mrm{g} y} \nonumber\\
    &\quad+ \int_{0}^{X_\mrm{T}} \mrm{d} x\, e^{V_\mrm{t}(x) + \delta_\mrm{g} x} \int_{-\infty}^{0} \mrm{d} y \, e^{-V_\mrm{t}(y) - \delta_\mrm{g} y} 
        +\int_{0}^{X_\mrm{T}} \mrm{d} x\, e^{V_\mrm{t}(x) + \delta_\mrm{g} x} \int_{0}^{x} \mrm{d} y \, e^{-V_\mrm{t}(y) - \delta_\mrm{g} y}\ ,
\end{align}
and substitute $x'=-x$ and $y'=-y$ in the second line and use the symmetry of the trapping potential, to obtain

\begin{align}
    \tau_\mrm{MFP}(X_\mrm{T}) &= \int_{-X_\mrm{T}}^{0} \mrm{d} x\, e^{V_\mrm{t}(x) + \delta_\mrm{g} x} \int_{-\infty}^{0} \mrm{d} y \, e^{-V_\mrm{t}(y)- \delta_\mrm{g} y}  
        - \int_{0}^{X_\mrm{T}} \mrm{d} x'\, e^{V_\mrm{t}(x') - \delta_\mrm{g} x'} \int_{0}^{x'} \mrm{d} y' \, e^{-V_\mrm{t}(y') + \delta_\mrm{g} y'} \nonumber\\
    &\quad + \int_{0}^{X_\mrm{T}} \mrm{d} x\, e^{V_\mrm{t}(x) + \delta_\mrm{g} x} \int_{-\infty}^{0} \mrm{d} y \, e^{-V_\mrm{t}(y) - \delta_\mrm{g} y} 
        + \int_{0}^{X_\mrm{T}} \mrm{d} x\, e^{V_\mrm{t}(x) + \delta_\mrm{g} x} \int_{0}^{x} \mrm{d} y \, e^{-V_\mrm{t}(y) - \delta_\mrm{g} y}\\
    &=\int_{-X_\mrm{T}}^{X_\mrm{T}} \mrm{d} x\, e^{V_\mrm{t}(x) + \delta_\mrm{g} x} \int_{-\infty}^{0} \mrm{d} y \, e^{-V_\mrm{t}(y) - \delta_\mrm{g} y} 
        + \int_{0}^{X_\mrm{T}} \mrm{d} x\, e^{V_\mrm{t}(x)} \int_{0}^{x} \mrm{d} y \, e^{-V_\mrm{t}(y)}\left[ e^{\delta_\mrm{g} (x-y)} - e^{-\delta_\mrm{g} (x-y)}\right]\ ,
\end{align}
\end{widetext}

where we re-labelled $x'$ and $y'$ to $x$ and $y$.
In general, $x-y \geq 0$ in the second double integral making it non-negative. Therefore, neglecting it yields a lower bound on the mean first-passage time. Also,
since $V_\mrm{t}(x) \geq V_\mrm{t}(0) = 0$, we find,
\begin{align}
    \tau_\mrm{MFP}(X_\mrm{T}) &\geq \int_{-X_\mrm{T}}^{X_\mrm{T}} \mrm{d} x\, e^{V_\mrm{t}(x) + \delta_\mrm{g} x} \int_{-\infty}^{0} \mrm{d} y \, e^{-V_\mrm{t}(y) - \delta_\mrm{g} y}\\
    &\geq \int_{-X_\mrm{T}}^{X_\mrm{T}} \mrm{d} x\, e^{\delta_\mrm{g} x} \int_{-\infty}^{0} \mrm{d} y \, e^{-V_\mrm{t}(y) - \delta_\mrm{g} y}\\
    &= \frac{2}{\delta_\mrm{g}} \sinh{\left(\delta_\mrm{g} X_\mrm{T}\right)}\,\int_{-\infty}^{0} \mrm{d} y \, e^{-V_\mrm{t}(y) - \delta_\mrm{g} y}\\
    &\geq 2 X_\mrm{T}\,\int_{-\infty}^{0} \mrm{d} y \, e^{-V_\mrm{t}(y) - \delta_\mrm{g} y}\,.
\end{align}
That is, for symmetric trapping potentials, the mean first-passage time exceeds its linear expansion for small $X_\mrm{T}$. Consequently, the power is maximized at vanishing threshold:
\begin{align}
    P(X_\mrm{T}) &=\nonumber \frac{2\delta_\mrm{g} X_\mrm{T}}{\tau_\mrm{MFP}(X_\mrm{T})} \leq P(X_\mrm{T} \rightarrow 0)\\
    &=  \delta_\mrm{g} \left[\int_{-\infty}^{0} \mrm{d} y \, e^{-V_\mrm{t}(y) - \delta_\mrm{g} y}\right]^{-1}\,.
\end{align}

\subsection{For symmetric potentials having a maximum stiffness, the harmonic shape is optimal}
In the preceding subsections, we have established that power is maximized for symmetric trapping potentials at vanishing ratcheting threshold. Here, we show that, when the stiffness of the potential is limited, a quadratic trap achieves maximum power.

We assume that the stiffness of the material is upper bounded by one. Since the trap has a single minimum fixed at $V(0)=0$, we can conclude that
\begin{align}
    V(x) \leq \frac{1}{2} x^2\,,
\end{align}
and, consequently, the extractable power at vanishing threshold reads
\begin{align}
    P(X_\mrm{T} \rightarrow 0) &=  \delta_\mrm{g} \left[\int_{-\infty}^{0} \mrm{d} y \, e^{-V_\mrm{t}(y) - \delta_\mrm{g} y}\right]^{-1}\\
    &\leq \delta_\mrm{g} \left[\int_{-\infty}^{0} \mrm{d} y \, e^{-y^2/2 - \delta_\mrm{g} y}\right]^{-1}\\
    &= \sqrt{\frac{2}{\pi}}\ \dgr \e^{-\dgr^2/2}\ \left[1 + \erf\left(\frac{\dgr}{\sqrt{2}}\right)\right]^{-1} \,,
\end{align}
i.e., when the stiffness is limited, the harmonic potential extracts maximum power.

\section{Limits on velocity and work extraction} 
\label{App:L}
We provide here more details on how to estimate the limits to the rate of work extraction and directed velocity. 
At sufficiently high trap stiffness, time scales in addition to the overdamped relaxation time $\trelo$ become important.
Fluid memory effects make relevant the time scale required for 
fluid vorticity to diffuse a single particle radius $r$,
\begin{align}
    \tau_{\mrm{f}} = \frac{r^{2}\rho}{\eta} \ , 
\end{align}
where $\eta$ the dynamical viscosity, and $\rho$ the density of the surrounding water. 
Solving for the trap stiffness where these two time scales are equivalent, $\trelo = \tau_{\mrm{f}}$, we obtain a critical trap stiffness
\begin{align}
    \kappa_{\mrm{f}}^{*} = \frac{6\pi\eta^2}{\rho r}\,,
\end{align}
where we recall that the Stokes dissipation for a spherical particle in an unbounded medium is $\gamma = 6\pi\eta r$.
Computing the velocity and power from \eqref{eq:vel_real} and \eqref{eq:P_real}, respectively, using the different values of $\kappa_{\mrm{f}}^{*}$ for the various bead diameters, we obtain rough estimates for the maximum velocity and output power
(Table~\ref{tbl:pow_vel_limits}).
Inertial effects also make relevant the time scale associated with the relaxation of momentum (velocity),
\begin{align}
    \tau_{v} = \frac{m}{\gamma} \,.
\end{align}
The ratio of these time scales is $\tau_v/\tau_\mrm{f} = (2/9) \Delta \rho/\rho \approx 0.2$ for silica beads in water, independent of particle size. Empirically, we have observed deviations from the single-time-scale overdamped theory when 
$\tau_r \ge 10 \tau_\mrm{f}$: in Fig.~\ref{fig:PvsBeadDiameter}, 
the red marker corresponding to a 5 \textmu m bead at $\kappa=200$ pN/\textmu m falls well below the predicted curve.
\begin{center}
    \begin{table}[!ht]
        \centering
        \begin{tabular}{|c|c|c|c|c|}
            \hline
            $d$\ [\textmu m] & $0.5$ & $1.5$ & $3$ & $5$ \\ \hline
            $\tau_{\mrm{f}}$\ [\textmu s] & $0.07$ & $0.6$ & $2.5$ & $7$  \\ \hline
            $\tau_{v}$ \ [\textmu s] & $0.01$ & $0.1$ & $0.6$ & $1.6$  \\ \hline
            $\kappa_{\mrm{f}}^{*}$ \ [\textmu N/m] & $60000$ & $20000$ & $10000$ & $6000$  \\ \hline
            $v^{*}$ \ [\textmu m/s] & $3000$ & $570$  & $200$  & $80$  \\ \hline
            $P^{*}\ \sr{\kT/s}$ & $470$ & $2400$  & $6800$ & $10000$ \\ 
            \hline
        \end{tabular}
        \caption{Fluid memory time scale, velocity relaxation time, critical trap stiffness and the associated velocity and output power (rows) for the different bead diameters (columns) that we consider.}
        \label{tbl:pow_vel_limits}
    \end{table}
\end{center}

\section{Error analysis} 
\label{App:M}
The error bars in all experimental figures represent the standard error of the mean. 
The relative uncertainties for the mean rate of stored gravitational energy $P$ differ from those for the trap $P_\mrm{trap}$, even though both quantities are calculated from the same position measurements. To understand why, we discuss how each quantity may be estimated from the underlying theoretical model. 

\begin{figure*}
    \centering
    \includegraphics[width = \linewidth]{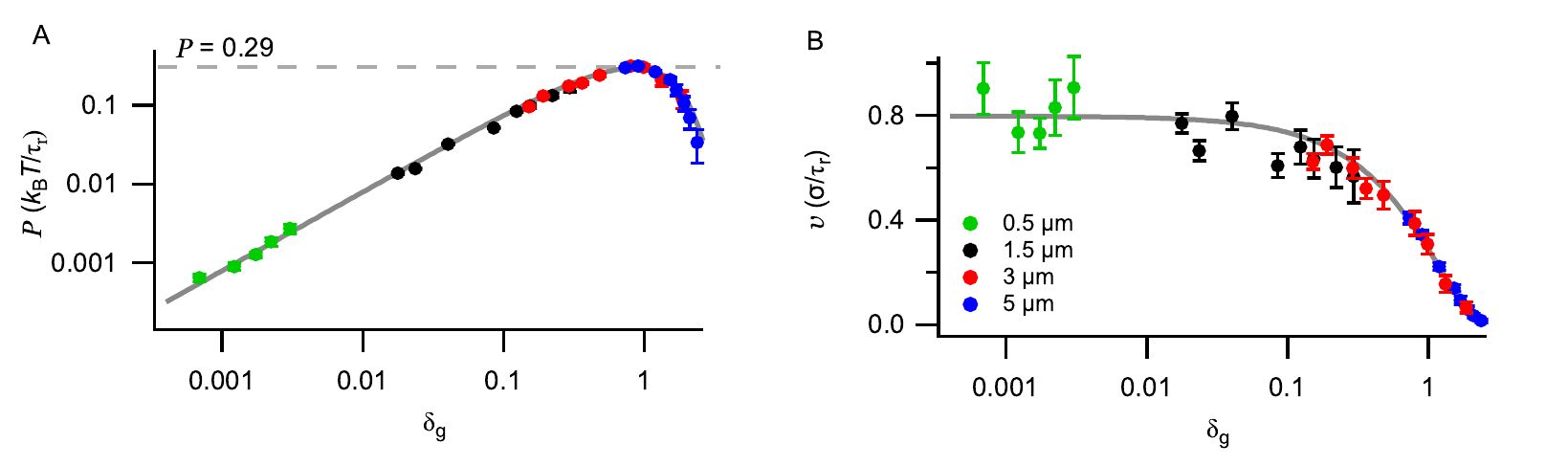}
    \caption{
        A) Scaled power and B) velocity as a function of $\dgr$ to show small $\dgr$ dependence. The symbols shows the same data as Fig. 4 (B) and (D) on log scale for scaled velocity and power, respectively.
        }
    \label{fig:Scaled_P_v_log}
\end{figure*}

The uncertainty in estimating  $P$ is calculated by propagating the errors in mean work extracted in $N_\mrm{traj}$ independent trajectories and in the $N_\mrm{traj}$ corresponding trajectory durations.  The mean power is defined as $P \equiv W^\mrm{traj} / t^\mrm{traj}$, where $W^\mrm{traj}$ and $t^\mrm{traj}$ are the mean extracted work and trajectory times, respectively. Note that although the trajectory lengths are fixed, typically, we exclude the first relaxation time of each trajectory, 
so
that the remainder of the trajectory has approximately steady-state statistics. This exclusion leads to fluctuations in the stored gravitational energy across
the trajectories, because the distance traveled in a fixed time is stochastic. Including the contribution from the stochastic first-passage time and neglecting the (small) covariance between work and time, we have
\begin{align}
    \left( \frac{\delta P}{P} \right)^2 
        = \left( \frac{\delta {W^\mrm{traj}}}{W^\mrm{traj}} \right)^2 +
        \left( \frac{\delta {t^\mrm{traj}}}{t^\mrm{traj}} \right)^2\,,
        \label{eq:P_err}
\end{align}
where $\delta W^\mrm{traj} = \sqrt{\mrm{Var}(W^\mrm{traj}) / N_\mrm{traj}}$, and $\mrm{Var}(W^\mrm{traj})$ is the variance of the gravitational energy stored in the trajectories. The error in the power $P_\mrm{trap}$ input by the trap is governed by a similar expression,
\begin{align}
    \left(\frac{\delta {P_\mrm{trap}}}{P_\mrm{trap}}\right)^2 = \left(\frac{\delta {W^\mrm{traj}_\mrm{trap}}}{W^\mrm{traj}_\mrm{trap}}\right)^2+
    \left(\frac{\delta {t^\mrm{traj}}}{t^\mrm{traj}}\right)^2\,.
    \label{eq:Ptrap_err}
\end{align}

In Eqs.~\ref{eq:P_err} and \ref{eq:Ptrap_err}, the uncertainties arising from trajectory-time fluctuations $\delta t^\mrm{traj}$ are identical.  As a result, we focus on estimating the work fluctuations in the first terms.  Since the actual values of these terms vary with parameters such as the trap stiffness $\kappa$, we focus on a typical case, $\delta_g = 0.8$, which maximizes $P$.

For gravitational energy, the fluctuation arises mainly from excluding the first relaxation time of the trajectory.  This leads to a position variance $\approx 2Dt_\mrm{r}$; hence, the variance of $W^\mrm{traj}$ is $\left( mg\,\sqrt{2Dt_\mrm{r}}  / k_\mrm{B}T \right)^2$. The mean work is $mg~X^\mrm{traj} / k_\mrm{B}T$, where $X^\mrm{traj}$ is the mean trajectory length. For $\delta_g = 0.8$, the predicted error estimates (mean $\pm$ standard error), in units of $ k_\mrm{B}T$, are $11.4 \pm 0.1$, which is very close to the experimental error estimates $W^\mrm{traj} = 11.3 \pm 0.1$.

For trap energy, the variance in $W_\mrm{trap}$ arises in part from feedback latency, the delay between measurement and response: If the response were instantaneous, then the only uncertainty would arise from measurement errors; however, the feedback delay $t_\mrm{s}$ allows the bead to relax.  By empirically tuning the feedback gain $\alpha$, we compensate for the mean shift, but fluctuations are still present. The variance in measured position at the reset time step is $2Dt_\mrm{s}+\sigma_m^2$, where $\sigma_m^2$ is the variance of position measurements. As the trap work depends on the square of the position, the variance of the square of the position is given by $2(2Dt_\mrm{s}+\sigma_m^2)^2$. Thus, for a trajectory with $N$ ratchet events, the trap-work variance is $\approx \left(\sqrt{2N}(2Dt_\mrm{s}+\sigma_m^2) \, \kappa/k_\mrm{B}T  \right)^2$. The predicted value of $\delta {W^\mrm{traj}_\mrm{trap}}$ for $N_\mrm{traj} = 81$, $N = 83$ and $\sigma_m = 1.7\, \mrm{nm}$ is $2.43\times10^{-2}$, which is very close to the experimental value, $2.47\times10^{-2}$.

The 0.5-\textmu m-bead experiments show larger relative fluctuations than the experiments using larger beads (Fig. \ref{fig:Scaled_P_v_log}). Measuring the position of this small bead size required a higher-NA objective and thus a smaller detection range. The trajectories then had half the length ($\approx 170$ nm) relative to the other experiments, with correspondingly fewer ratchet events.  Here we focus on errors in power in the limit of small $\delta_g$, where the directed velocity is maximized.  For $\delta_g = 7.5\times10^{-4}$, the experimental mean work $W^\mrm{traj} = 3.4 \pm 0.3 \times 10^{-2}$, which is comparable to the expected values of $3.3 \pm 0. 1\times 10^{-2}$, but with higher uncertainty. For trap energies, the experimental error estimates are $\delta {W^\mrm{traj}_\mrm{trap}} = 3.8$, and the predicted values are of the order $\delta {W^\mrm{traj}_\mrm{trap}} = 0.95$. The position measurements were significantly affected by the measurement noise ($\sigma_m \approx 4 $ nm) as it is comparable to the standard deviation of the bead in the trap ($\sigma \approx 5$ nm).

The experimental variances of the energies
are larger than predicted, because of vibrations and low-frequency drift in the mechanical setup and because of high-frequency intensity fluctuations in the detection laser. From the above two case studies, we see that the uncertainties, both absolute and relative, change considerably with $\delta_g$.  Nonetheless, they follow the trend that $\delta {P_\mrm{trap}}/P_\mrm{trap} > \delta {P}/P$. The reason is that although both energies are calculated from the bead positions, the uncertainty in $P$ arises from the ``single'' measurement uncertainty at the beginning of the trajectory, whereas $P_\mrm{trap}$ arises from the accumulated errors in the $N$ ratchet events in a trajectory.

\bibliographystyle{apsrev4-2}

\end{document}